\documentclass[useAMS,usenatbib]{mn2e}
\usepackage{times}
\usepackage{graphicx}
\usepackage{aas_macros}
\usepackage[usenames]{color}
\definecolor{New}{rgb}{0,0,1}
\definecolor{Todo}{rgb}{1,0,1}

\title[ALS 1135 in the OB association Bochum 7]{The massive eclipsing system ALS 1135 and variable stars in the field of the distant OB association Bochum 7}
\author[G. Michalska et al.]{G. Michalska$^{1,2}$, E. Niemczura$^{1,3}$, A. Pigulski$^1$, M. St\c{e}\'slicki$^{1,4}\thanks{Visiting Scientist}$, A. Williams$^5$\\
$^1$Instytut Astronomiczny, Uniwersytet Wroc{\l}awski, Kopernika 11, 51-622 Wroc{\l}aw, Poland\\
$^2$Universidad de Concepci\'on, Departamento de Astronom\'{\i}a, Casilla 160-C, Concepci\'on, Chile\\
$^3$Astronomick\'y \'ustav AV \v{C}R, Fri\v{c}ova 298, 251 65 Ond\v{r}ejov, Czech Republic\\ 
$^4$High Altitude Observatory, NCAR, P.O. Box 3000, Boulder, CO 80307, USA\\
$^5$Perth Observatory, Walnut Road, Bickley, Perth 6076, Australia\\
}
\begin{document}

\date{Accepted Received ; in original form É}

\pagerange{\pageref{firstpage}--\pageref{lastpage}} \pubyear{2012}

\maketitle

\label{firstpage}

\begin{abstract}
Using photometric and spectroscopic observations of the double-lined early-type eclipsing binary system ALS\,1135, a member of the distant OB association Bochum 7, we derived the new physical and orbital parameters of its components. The masses of both components were derived with an accuracy better than 1 per cent, their radii, with an accuracy better than 3 per cent. Since the primary's mass is equal to about 25~$M_\odot$, its radius was subsequently used to derive the age of the system which is equal to 4.3 $\pm$ 0.5~Myr. The result shows that this method represents a viable alternative to isochrone fitting.

A photometric search of the field of ALS\,1135 resulted in the discovery of 17 variable stars, including seven pulsating ones. One of them is an SPB star belonging to Vel OB1, the other six are $\delta$ Scuti stars. Of the six $\delta$~Scuti stars three might belong to Vel OB1, the other two are likely members of Bochum 7. Given the age of Bochum 7, these two stars are probably pre-main sequence pulsators. In addition, we provide $UBVI_{\rm C}$ photometry for about 600 stars in the observed field.
\end{abstract}


\section{Introduction}
It is well known that a combination of the light curve of an eclipsing binary and its double-lined spectroscopic orbit allows to derive directly masses and radii of the components of a binary system. Recently, \citet*{torr2010} presented a list of 95 detached systems with masses and radii of both components known with an accuracy of 3\% or better.  It is remarkable that out of 190 components of these systems, only three have masses greater than 20\,$M_{\odot}$. On the other hand, good knowledge of masses for such stars is very important for understanding the progenitors of the core-collapse supernovae, mass loss, fast rotation and phenomena related to stellar formation. 
In addition, accurate determination of the parameters of massive stars \citep*[see, e.g.,][]{hild1996} allows precise determination of their ages. Since massive binaries are often the members of young clusters or associations, this opens a possibility of an independent derivation of the age of the parent stellar system. This is important because for young stellar systems most methods of age determination like isochrone fitting provide only a very rough estimate. In the present paper, we shall derive the age of the OB association Bochum 7 using the parameters of its member, ALS\,1135.

ALS\,1135 $=$ CPD\,$-$45$^\circ$2920 ($\alpha_{\rm 2000} =$ 08$^{\rm h}$43$^{\rm m}$49.8$^{\rm s}$, $\delta_{\rm 2000}$ $=$ $-$46$^\circ$07$^\prime$09$^{\prime\prime}$) was classified as an OB star in the catalog of Luminous Stars in the Southern Milky Way \citep{stsa1971}. The first MK classification of the star, O6\,III, was given by \citet{vidr1993}. \citet*{cort1999} and \citet*{cort2003} discovered that ALS\,1135 is a single-lined spectroscopic binary with a period of 2.7532\,d, and classified the main component as O6.5\,V((f)). Furthermore, the star was found to be an eclipsing system from the ASAS photometry \citep{pojm2003}. Taking these results into account, \citet[][hereafter FLN2006]{feni2006} re-examined spectra used in previous investigations finding faint He\,I lines of the secondary component. 
From radial velocities and the ASAS photometry, FLN2006 obtained orbital solution and physical parameters of both components. The spectral type of the secondary component was estimated as B1\,V. Later on, in the same ASAS photometry of ALS\,1135, \citet{pimi2007} found additional periodic variations with a frequency of 2.31095~d$^{-1}$. The possibility that these variations are due to pulsations of the primary component prompted us to carry out a follow-up photometric and spectroscopic campaign devoted to this star. Results of this campaign are reported in the present paper; a by-product is the discovery of 17 variable stars in the vicinity of ALS\,1135.

\begin{figure*}
\centering
\includegraphics[width=9.5cm]{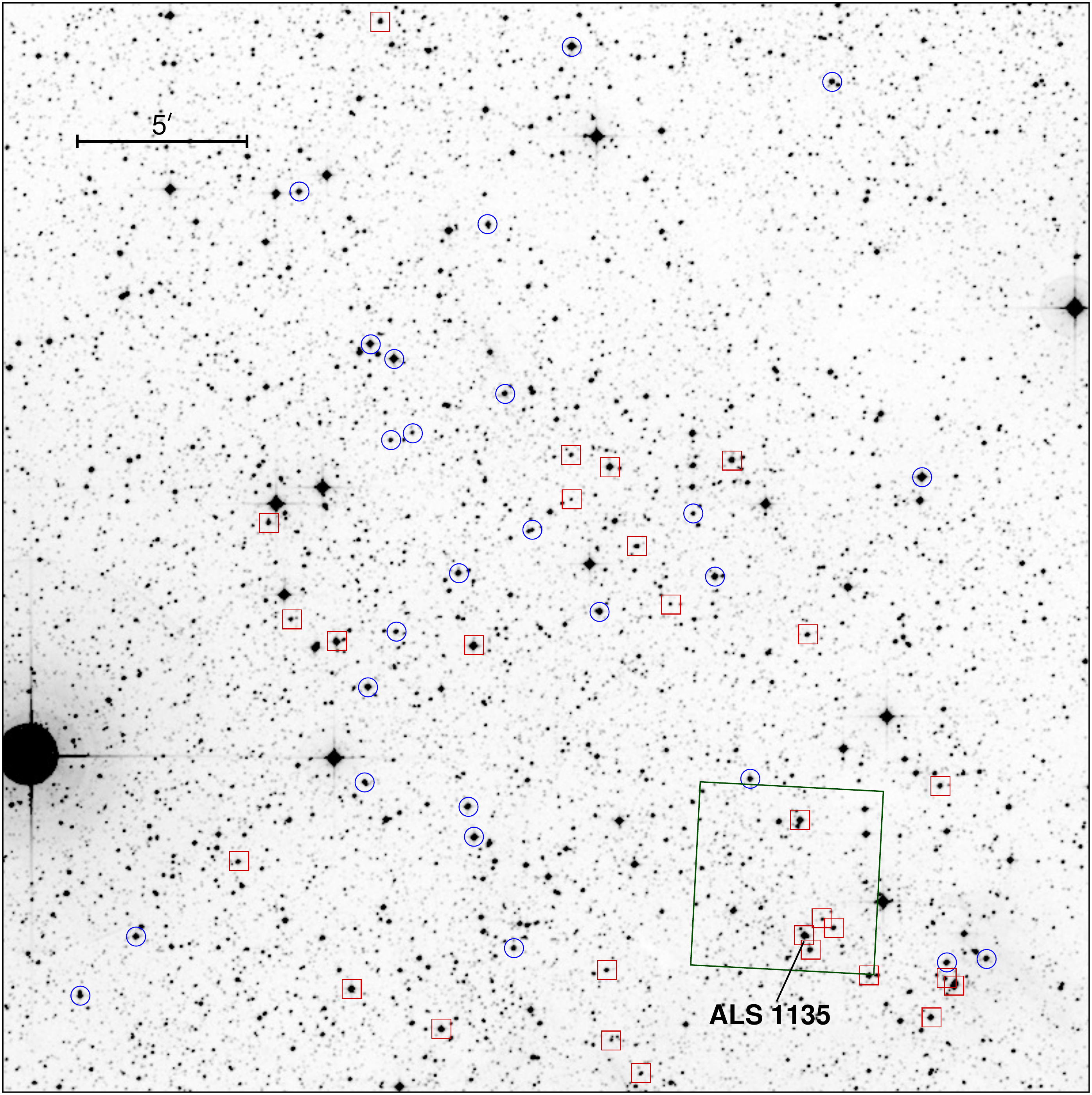}
\caption{A 32$^{\prime}\times$ 32$^{\prime}$ fragment of the DSS2-Red plate centred at ($\alpha_{\rm 2000.0}$, $\delta_{\rm 2000.0}$) $=$ (8$^{\rm h}$44$^{\rm m}$35{\fs}5, $-$45$^\circ$56$^\prime$) containing the observed field around ALS\,1135 (large square). Open circles and squares represent stars belonging to Vel OB1 and Bochum 7 (Vel OB3), respectively, identified using the photometry and spectroscopy of \citet*{cort2007}. The bright star on the left is HR 3487 (a Vel).}
\label{map}
\end{figure*}

The early-type binary ALS\,1135 is situated in a very interesting part of the Milky Way populated by young OB associations. Although the presence of at least five different associations was proposed in this area of the sky \citep{egge1982,bass1982,slre1988,turn1993,tovm1993,sung1999}, the reality of only three of them seems to be relatively well established. These are: the nearest Vel OB2 \citep{slre1988,deze1999} located at a distance of about 0.4~kpc, Vel OB1 at a distance of 1.5--1.9~kpc \citep{hump1978,bass1982,slre1988}, and Bochum 7 (Vel OB3) at a distance of 4--6~kpc \citep{mill1972,movo1975,slre1988,reed2000}. 
The latter association is known to harbour the Wolf-Rayet star ALS\,1145 $=$ WR\,12 \citep{movo1975,sung1999} and at least several OB stars including ALS\,1135 \citep{movo1975}. The membership of ALS\,1135 in Bochum 7 was concluded from its spectroscopic parallax \citep{movo1975,crow2006}. It was also confirmed by a direct comparison of the radial velocity of the star with that of other members of Bochum 7 \citep{cort2003}.  Fig.\,\ref{map} shows the position of ALS\,1135 and the members of Vel OB1 and Bochum 7 in its vicinity.

The paper is organised as follows. In Sect.\,2, we describe spectroscopic and photometric observations and data reduction methods. An analysis of $UBVI_{\rm C}$ light curve and radial-velocity curve of ALS\,1135 is presented in Sect.\,3. In that section, we derive new parameters of the system. Variable stars found in the field we observed are discussed in Sect.\,4. Then, in Sect.~5, photometric diagrams for the observed field are studied. Finally, a summary and conclusions are given in Sect.\,6. 

\section{Observations and Reductions}
\subsection{Photometry}
The $UBVI_{\rm C}$ observations of the eclipsing system ALS\,1135 were carried out at South African Astronomical Observatory (SAAO) during 13 nights between January 23 and February 5, 2008. We used the SAAO's 1-m Cassegrain telescope equipped with a 1024\,$\times$\,1024 CCD camera covering an area of about 5.3$^{\prime}\times$\,5.3$^\prime$ (Fig.\,\ref{map}). In total, about 450 frames in $U$, 880 in $B$, 1700 in $V$ and 1900 in an $I_{\rm C}$ filter were taken. The exposure times ranged from 15 to 120 s, depending on the filter, seeing and sky transparency.

Additionally, $BVI_{\rm C}$ observations were taken with the 0.6-m Perth-Lowell Automated Telescope at Perth Observatory, Australia. The data were collected between November 2007 and January 2009. Most of the observations (more than 90\%), however, were taken between January 21 and 29, 2008. Using this telescope, we obtained about 780, 860 and 850 frames in $B$, $V$ and $I_{\rm C}$, respectively. All frames were calibrated in a standard way and reduced with the DAOPHOT II package \citep{stet1987}. 

\subsection{Spectroscopy}
Spectroscopic observations of ALS\,1135 were carried out between January 25 and 27, 2008 with the ESO New Technology Telescope (NTT, La Silla, Chile) and the ESO Multi-Mode Instrument (EMMI) in the cross-dispersed echelle mode. The EMMI {\'e}chelle spectroscopy was done in the REMD (Red Medium Dispersion Spectroscopy) mode in the wavelength range from about 4800 to 6800~{\AA} and a resolving power equal to 35\,000. We used grating \#10 and 1$^{\prime\prime}$-wide slit. In total, 48 spectra with 15-min integrations were obtained. 

\begin{table*}
 \centering
 \begin{minipage}{140mm}
  \caption{\label{trv}Heliocentric radial velocities of both components of ALS\,1135.}
  \begin{tabular}{@{}crrrrccrr}
  \hline
&\multicolumn{2}{c}{Primary}&\multicolumn{2}{c}{Secondary}&&&\multicolumn{2}{c}{Primary}\\
HJD  & \multicolumn{1}{c}{HRV} & res$_{\rm HRV}^\star$ & HRV &  res$_{\rm HRV}^\star$ &\quad & HJD &\multicolumn{1}{c}{HRV} & res$_{\rm HRV}^\star$ \\
2454400.+ & [km\,s$^{-1}$] & [km\,s$^{-1}$]   & [km\,s$^{-1}$] & [km\,s$^{-1}$] &&2454400.+ & [km\,s$^{-1}$] & [km\,s$^{-1}$] \\
\hline\noalign{\smallskip}  
  90.60405 &$-$29.9 &  +4.7  & +352 &   +1   &&  92.57312 & +152.6 & $-$0.1\\
  90.61539 &$-$31.2 &  +5.1  & +340 &$-$16 &&  92.58446 & +147.3 & $-$3.5\\
  90.63777 &$-$39.4 &  +0.2  & +352 &$-$15 &&  92.59584 & +144.3 & $-$4.5\\
  90.64909 &$-$35.8 &  +5.3  & +389 &  +17   &&  92.60717 & +146.4 & $-$0.3\\
  90.66042 &$-$35.6 &  +6.9  & +382 &   +6  &&  92.62726 & +141.1 & $-$1.7\\
  90.67176 &$-$40.6 &  +3.3  & +366 &$-$15 &&  92.63859 & +145.0 & +4.4\\
  90.69420 &$-$46.9 & $-$0.5  & +394 &   +5  &&  92.64992 & +139.8 & +1.5\\
  90.70553 &$-$43.5 &  +4.1  & +398 &   +5  &&  92.66124 & +136.3 & +0.4\\
  90.71686 &$-$48.2 &  +0.5  &     &             &&  92.68136 & +133.1 & +1.5\\
  90.72818 &$-$52.4 &$-$2.7 &     &             &&  92.69270 & +129.3 & +0.3\\
  90.75005 &$-$53.0 &$-$1.5 & +410 &  +4 &&  92.70407 & +127.9 & +1.4\\
  90.76139 &$-$51.2 & +1.1 & +367 &$-$42 &&  92.71542 & +120.8 & $-$3.0\\
  90.77287 &$-$64.2 &$-$11.1&     &             &&  92.73953 & +121.3 & +2.9\\
  90.78422 &$-$61.4 &$-$7.7 & +432 &  +19   &&  92.75090 & +114.9 & $-$1.6\\
  90.80525 &$-$61.4 &$-$6.7 &     &             &&  92.76226 & +111.5 & $-$3.3\\
  90.81658 &$-$65.0 &$-$9.8 & +406 &$-$13 &&  92.77363 & +110.7 & $-$2.6\\
  90.82792 &$-$65.9 &$-$10.3&     &             &&  92.79434 & +108.5 &$-$2.3\\
  90.83925 &$-$52.7 & +3.2  &     &             &&  92.80573 & +106.2 & $-$3.1\\
  90.86219 &$-$59.8 &$-$3.5 & +425 &  +3    &&  92.81712 & +101.3 &$-$6.3 \\
  90.87410 &$-$51.4 & +5.0 & +384 &$-$39 &&  92.82849 &   +96.8 & $-$8.9\\
  91.84062 &  +137.0 & +8.5 &     &             &&  92.84918 &    +94.7 & $-$6.9\\
  91.85195 &  +146.6 &+15.7 &     &              &&  92.86055 &    +98.7 & $-$0.1\\
  91.86328 &  +148.3 &+14.9 &     &              &&  92.87192 &    +91.4 & $-$3.9\\
  91.87460 &  +136.4 &+0.7&     &              &&  92.88329 &    +91.4 & +0.3\\
\noalign{\vskip1pt}\hline\noalign{\vskip1pt}
\multicolumn{8}{l}{$^\star$ residuals from the orbital solution obtained with the W-D program (see Sect.\,\ref{lcwd}).}\\
\end{tabular}
\end{minipage}
\end{table*}

Moreover, a single spectrum of ALS\,1135 was obtained with the Magellan Inamori Kyocera {\'E}chelle (MIKE) spectrograph \citep{bern2003} attached to the 6.5-m Magellan-Clay Telescope at Las Campanas Observatory, Chile. The observation was taken on January 5, 2008, with a 0.7$^{\prime\prime}$-wide slit, which resulted in a resolving power of 41\,000 and 32\,000 in the blue and red part of the spectrum, respectively. The spectrum covered wavelengths from 3600 to 7600\,{\AA}, with the signal-to-noise ratio of about 100. 

In order to reduce the spectra, standard IRAF\footnote{IRAF (Image Reduction and Analysis Facility) is distributed by the National Optical Astronomy Observatories, which are operated by the Association of Universities for Research in Astronomy, Inc., under cooperative agreement with the National Science Foundation.} procedures were used. The raw data were bias subtracted, corrected for pixel-to-pixel variations (flat-field) and sky subtracted. Wavelength calibrations were carried out using Th-Ar lamps. Next, the individual rows of each spectrum were normalised by fitting a blaze function. The rows were subsequently combined into a single spectrum.

\section{ALS 1135}
\label{alss}
Recently, \citet{pimi2007} have presented results of a search for pulsating components in eclipsing binary systems using the ASAS-3 database. They found out-of-eclipse periodic variability in eleven systems, including ALS\,1135 (ASAS\,084350$-$4607.2). In addition to variation caused by binarity, ALS\,1135 showed sinusoidal variations with a period equal to 0.4327\,d. Since p-mode pulsations in O-type stars are allowed by the theory \citep{pamy1999}, \citet{pimi2007} suspected that the periodicity could indicate a $\beta$\,Cephei-type pulsations in the O-type primary. They pointed out, however, that contamination of the ASAS photometry due to the low spatial resolution is an alternative. 
Indeed, our analysis of good-quality photometric data showed that variations with the 0.4327\,d period originate in the nearby EW-type eclipsing system V4 (see Sect.\,\ref{secl}) which was not resolved in the ASAS data. Thus, the system can be studied without complications caused by pulsations. The accurate orbital and physical parameters of such systems are of great importance and can be used for testing theoretical stellar models \citep[see, e.g.,][]{bona2009}. In particular, in the present paper, we use the parameters of the ALS\,1135 to derive its age (see Sect.\,\ref{age}).

\subsection{\label{rvap}Radial velocities and atmospheric parameters}
In order to obtain atmospheric parameters of the primary component we compared the high-resolution MIKE spectrum with theoretical spectra from the TLU\-STY\-/SYNSPEC codes given in the OSTAR2002 grid \citep{lahu2003}. The OSTAR2002 grid consists of about 700 metal line-blanketed, non-LTE, plane-parallel, hydrostatic model atmospheres. The synthetic spectra were calculated for effective temperatures between 27\,500 and 55\,000\,K with a step equal to 2500\,K and surface gravities ranging from 3.0 to 4.75\,dex with a step equal to 0.25\,dex. In our fitting we adopted fluxes calculated for solar metallicity and assumed microturbulence velocity equal to 10\,km\,s$^{-1}$. The model spectra were convolved with rotational and instrumental profiles.

\begin{figure}
\centering
\includegraphics[width=70mm]{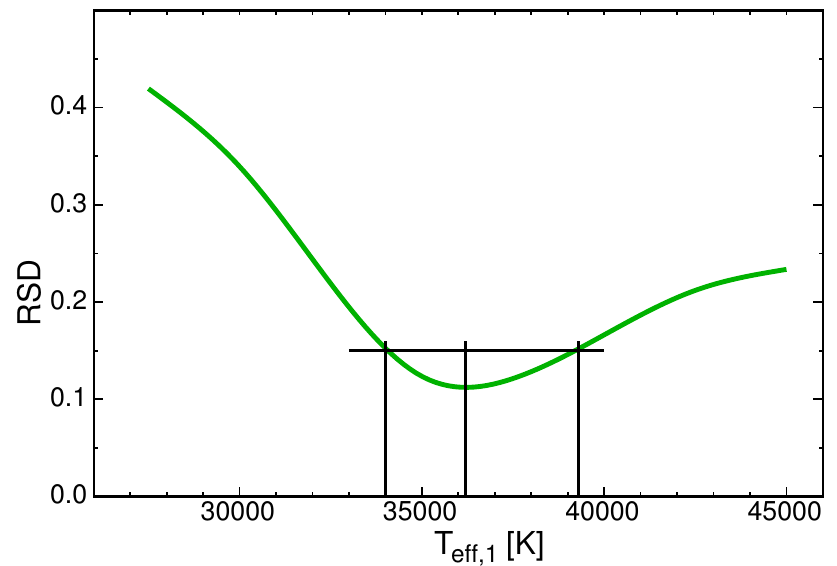}
\caption{Residual standard deviation (RSD) of the fit of atmospheric models to the observed spectra in the vicinity of twelve helium lines. They are plotted as a function of the assumed effective temperature of the primary, $T_{\rm eff,1}$. For all models, $\log (g_1/{\rm cm\,s}^{-2})=$ 3.75~dex was adopted. See text for details.} 
\label{rsd}
\end{figure}

Despite the fact that the MIKE spectrum was taken close to the primary eclipse, the contribution of the secondary to the total flux at that epoch can be neglected in the procedure of obtaining stellar parameters. In other words, stellar parameters derived from the profiles of stellar lines in this spectrum can be attributed to the primary. The determination of primary's surface gravity, $\log g_1$, was based on the H$\gamma$ and H$\beta$ line profiles and resulted in $\log (g_1/{\rm cm\,s}^{-2})=$ 3.75. In order to derive effective temperature of the system's primary, $T_{\rm eff,1}$, we used He\,I 4026\,{\AA}, 4145\,{\AA}, 4471\,{\AA}, 4713\,{\AA}, 4921\,{\AA}, 5061\,{\AA}, 5871\,{\AA}, He\,II 4201\,{\AA}, 4388\,{\AA}, 4542\,{\AA}, 4686\,{\AA} and 5411\,{\AA} as diagnostic lines. The best fit of the model fluxes to the observed lines was obtained as a result of minimising residual standard deviation (RSD) of the fit. 
The RSD dependence on $T_{\rm eff,1}$ is clearly asymmetric (Fig.~\ref{rsd}). Adopting 2500~K as a reasonable uncertainty of $T_{\rm eff,1}$ for a mid-O type star, we get $T_{\rm eff,1} =$ 36\,200$^{+3100}_{-2200}$ K, if we demand that the upper and lower value of $T_{\rm eff,1}$ should correspond to the same value of the RSD (horizontal line in Fig.~\ref{rsd}). Additionally, the projected rotational velocity, $V \sin i$, was found to be equal to 195\,$\pm$\,15\,km\,s$^{-1}$. FLN2006 assumed the effective temperature of the primary component equal to 37\,870\,K using spectral type vs.~effective temperature calibration of \citet*{mart2005}. 

The EMMI/NTT data were used to derive heliocentric radial velocities (HRV) of the components of ALS 1135. For the primary component, radial velocities were derived by fitting a rotationally-broadened synthetic spectrum from the above-mentioned grid with $\log (g_1/{\rm cm\,s}^{-2}) =$ 3.75 dex and $T_{\rm eff,1}=$ 37\,500\,K, the closest to the best-fit one. In order to avoid the influence of telluric and interstellar lines, we fitted the fluxes only in the vicinity of seven spectral lines. These lines were the following: He\,II 5411\,{\AA}, O\,III 5592\,{\AA}, C\,IV 5801 and 5812\,{\AA}, He\,I 5876\,{\AA}, He\,II 6406\,{\AA} and He\,II 6527\,{\AA}. 
Both synthetic and observed spectra were normalised and rebinned in $\ln\lambda$ before fitting. For the secondary component we were able to derive radial velocities only from a single line, He\,I 5876\,{\AA}, and only from the best spectra taken on the night of January 25, 2008, close to quadrature. At this phase, the lines of components were well separated allowing determination of the HRV for the secondary by fitting the line with a Gauss function. The values of the HRV we derived are given in Table \ref{trv}.

\begin{figure}
\centering
\includegraphics[width=84mm]{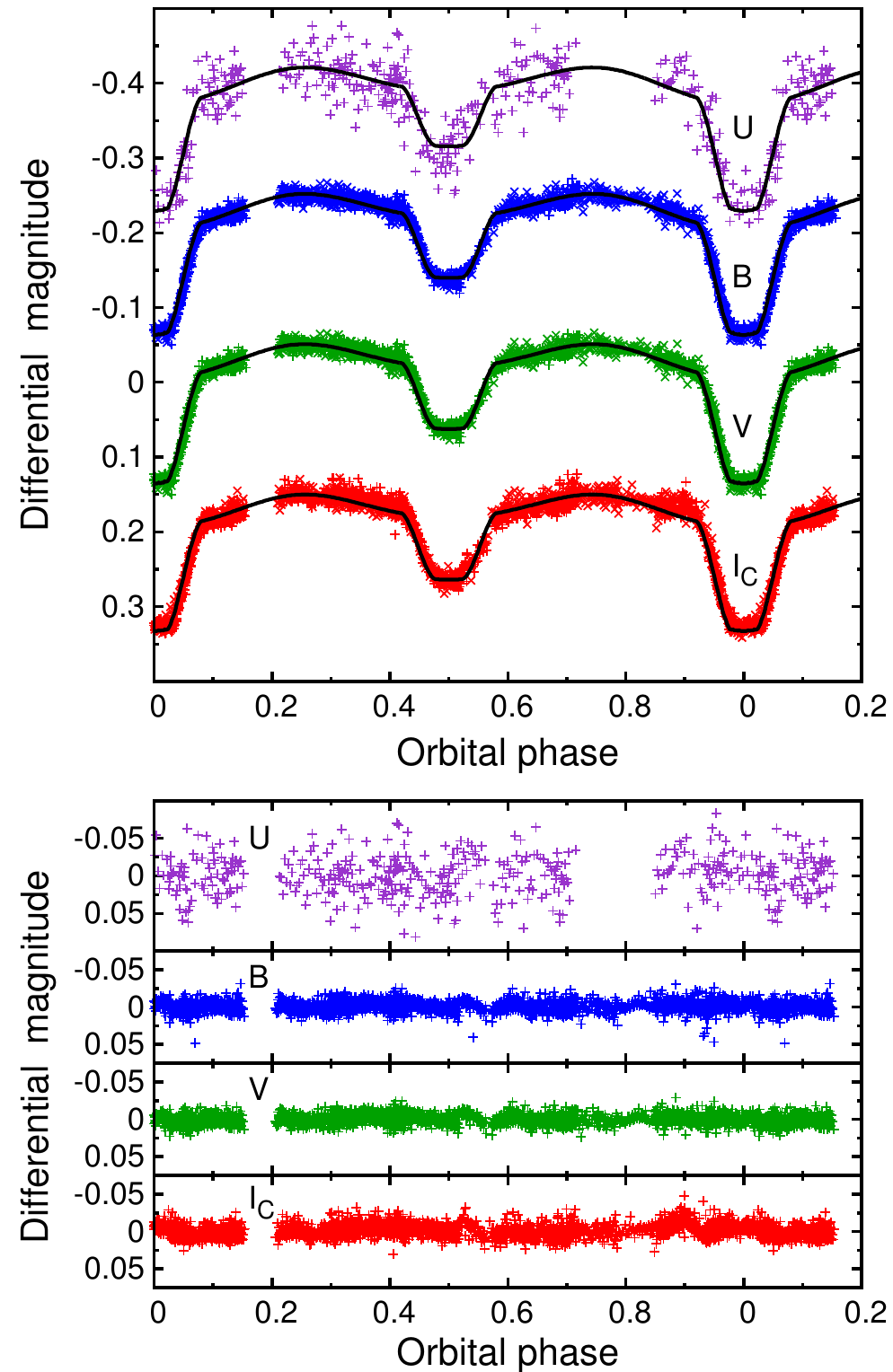}
\caption{{\it Top:} Phase diagrams of the $UBVI_{\rm C}$ observations of the eclipsing system ALS 1135. Offsets were applied in order to separate light curves in different bands. The continuous lines show the best fit obtained by means of the W-D program. The parameters of this fit are given in Table \ref{talspar}. {\it Bottom:} Residuals from the fit.} 
\label{lc1}
\end{figure}

\begin{figure}
\centering
\includegraphics[width=84mm]{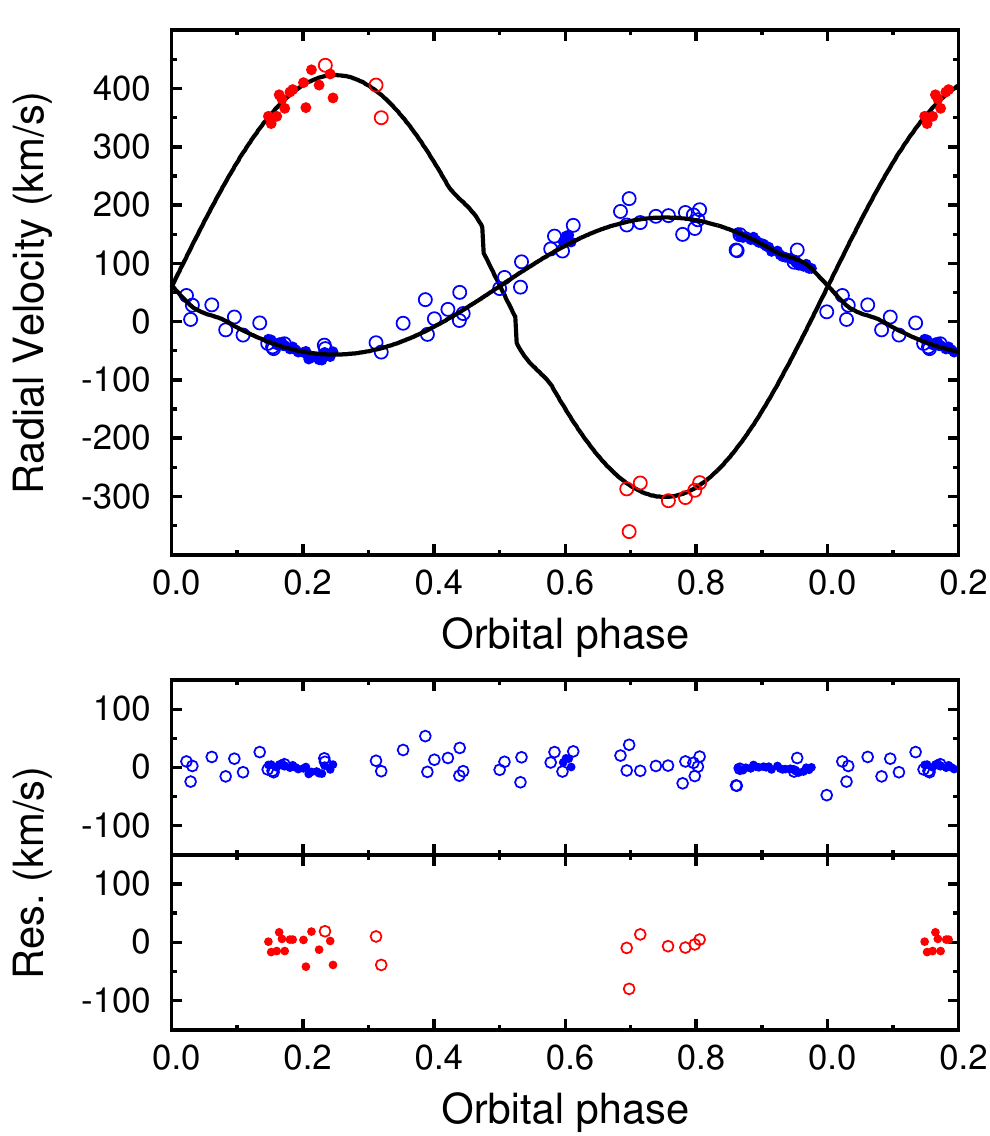}
\caption{{\it Top:} Radial velocities of both components. The radial velocities derived in this paper are shown with filled circles, those obtained by FLN2006 with open circles. The solid lines represent the best fit obtained by means of the W-D program. The parameters of this fit are given in Table \ref{talspar}. {\it Bottom:} Residuals from the fit.} 
\label{rv1}
\end{figure}

\subsection{Modelling light and radial velocity curves}
\label{lcwd}
The physical parameters of the components of ALS\,1135 were determined using the newest version of the Wilson-Devinney (hereafter W-D) program \citep{wide1971,wils1979,wils1990,vhwi2007}. Our $UBVI_{\rm C}$ and radial velocity curves were modelled simultaneously. Fitting radial velocity curves, we combined FLN2006 and our data. Due to the larger uncertainties of the former, they were assigned lower weights.

First, we determined the orbital period and the time of the primary minimum. Combining ASAS-3 and our $V$-band data, we derived the following ephemeris:
\begin{equation}
{\rm Min\,I} = \mbox{HJD\,2452070.144(13)}+ \mbox{2{\fd}753189(14)}\times E,
\end{equation}
where $E$ is the number of elapsed cycles. The numbers in parentheses denote the r.m.s.~errors of the preceding quantities with the leading zeroes omitted. The phase diagrams in $U$, $B$, $V$ and $I_{\rm C}$ filters are shown in Fig.\,\ref{lc1}.

The W-D program was run with detached geometry (MODE $=$ 2). We used detailed reflection model with two reflections (MREF\,$=$\,2, NREF $=$ 2). Bolometric albedos and gravity darkening coefficients equal to 1.0 were assumed as for stars with radiative envelopes. The coefficients of the logarithmic limb darkening law were computed for the effective temperature and surface gravity from the formula given by \citet{vanh1993}. Additionally, we assumed that the components rotate synchronously and have no spots. Since both eclipses have equal widths and the secondary minimum occurs at phase 0.5, we adopted circular orbit. We obtained three solutions for three assumed $T_{\rm eff,1}$, 34\,000, 36\,200, and 39\,300~K, in accordance with the spectroscopic determination of $T_{\rm eff,1}$ in Sect.~3.1. 
At the beginning, the mass ratio ($q$) and the radial velocity of the system's barycentre ($V_{\gamma}$) were derived by fitting sine functions to the phased radial-velocity curves. Keeping these parameters fixed in the W-D program, we adjusted phase shift ($\varphi$), surface potentials ($\Omega_1$ and $\Omega_2$), effective temperature of the secondary component ($T_{\rm eff,2}$), inclination ($i$) and monochromatic luminosity of the primary component ($L_1$). The iterations were repeated until the solution converged. Then, the mass ratio, the orbital semi-major axis ($a$) and radial velocity of barycentre were also freed in the fitting procedure and all nine parameters were adjusted simultaneously. 
The W-D program provides also masses, radii, surface gravities of both components and the luminosity of the secondary component. The solutions for the three different $T_{\rm eff,1}$ led obviously to different values of $T_{\rm eff,2}$. The remaining parameters remained, however, virtually the same. We therefore decided to provide a single solution (Table \ref{talspar}) incorporating the differences due to the allowed range of $T_{\rm eff,1}$ into the uncertainties of parameters. The exceptions are for obvious reasons $T_{\rm eff,2}$, $\log L/L_\odot$ and $M_{\rm bol}$; for these three parameters we provide values corresponding to the three assumed values of $T_{\rm eff,1}$.

The full set of parameters is given in Table \ref{talspar}. The final fit to the light and radial velocity curves is shown in Fig.\,\ref{lc1} and \ref{rv1}, respectively. The standard deviations of the residuals for the SAAO data are equal to 0.027, 0.006, 0.006 and 0.008~mag in $U$, $B$, $V$ and $I_{\rm C}$, respectively. For the data from Perth, the standard deviations are similar and equal to 0.007, 0.007 and 0.008~mag for $B$, $V$ and $I_{\rm C}$, respectively. The standard deviations of the residuals of our radial velocity data are equal to 5.4\,km\,s$^{-1}$ and 18.6\,km\,s$^{-1}$ for the primary and secondary component, respectively. In the case of the FLN2006 the residuals amount to 19.7\,km\,s$^{-1}$ for the primary and 29.6\,km\,s$^{-1}$ for the secondary.

\begin{table}
 \centering
 \begin{minipage}{85mm}
  \caption{\label{talspar}The orbital and physical parameters of ALS 1135 derived by means of the W-D program.}
  \begin{tabular}{@{}ccc}
  \hline
Parameter  & \multicolumn{1}{c}{Primary (1)} &\multicolumn{1}{c}{Secondary (2)} \\
\noalign{\vskip1pt}\hline\noalign{\vskip1pt} 
$\varphi$ & \multicolumn{2}{c}{$+$0.0008 $\pm$ 0.0002} \\ 
$T_{\rm eff}$ [K] 
& $\left\{\begin{array}{l}\mbox{39\,300}^\star\\
                                     \mbox{36\,200}^\star\\
                                     \mbox{34\,000}^\star\\
\end{array}\right. $
& $\begin{array}{l}\mbox{27\,120} \pm \mbox{80}\\
                            \mbox{25\,260} \pm \mbox{80}\\
                            \mbox{23\,890} \pm \mbox{70}\\
      \end{array} $\\
$\Omega$ & 3.060 $\pm$ 0.007 & 3.826 $\pm$ 0.020 \\
$q=M_2/M_1$ & \multicolumn{2}{c}{0.326 $\pm$ 0.003} \\
$i$\,[$^{\circ}$] & \multicolumn{2}{c}{79.6 $\pm$ 0.2} \\
$a$\,[$R_{\odot}$] & \multicolumn{2}{c}{26.67 $\pm$ 0.22} \\
$V$$_{\gamma}$\,[km\,s$^{-1}$] & \multicolumn{2}{c}{$+$61.2 $\pm$ 0.3} \\
$K$\,[km\,s$^{-1}$] & 118.5 $\pm$ 1.0 & 363.2 $\pm$ 3.9\\
$[L_{1,2}/(L_1+L_2)]_{\rm U}$ & 0.9521 $\pm$ 0.0033 & 0.0479 $\pm$ 0.0033\\
$[L_{1,2}/(L_1+L_2)]_{\rm B}$ & 0.9442 $\pm$ 0.0012 & 0.0558 $\pm$ 0.0012\\
$[L_{1,2}/(L_1+L_2)]_{\rm V}$ & 0.9420 $\pm$ 0.0009 & 0.0580 $\pm$ 0.0009\\
$[L_{1,2}/(L_1+L_2)]_{\rm I}$ & 0.9401 $\pm$ 0.0008 & 0.0599 $\pm$ 0.0008\\
$R\mbox{ (pole)}/a$ &0.3623 $\pm$ 0.0016  & 0.1302 $\pm$ 0.0010 \\
$R\mbox{ (point)}/a$ &0.3912 $\pm$ 0.0022  & 0.1323 $\pm$ 0.0011 \\
$R\mbox{ (side)}/a$ &0.3748 $\pm$ 0.0018 & 0.1308 $\pm$ 0.0010 \\
$R\mbox{ (back)}/a$ &0.3835 $\pm$ 0.0020  & 0.1320 $\pm$ 0.0011\\
$R$\,[$R_{\odot}$] & 9.99 $\pm$ 0.09 & 3.49 $\pm$ 0.03\\
$M$\,[$M_{\odot}$] & 25.3 $\pm$ 0.7 & 8.25 $\pm$ 0.17\\
$\log (g/{\rm cm}\,{\rm s}^{-2}$) & 3.842 $\pm$ 0.016 & 4.268 $\pm$ 0.017\\
$V\sin i$\,$^{\star\star}$ [km s$^{-1}$] & 180.5 $\pm$ 2.3 & 63.1 $\pm$ 1.0\\
$\log(L/L_\odot)$
& $\left\{\begin{array}{l}\mbox{5.330} \pm \mbox{0.011}\\
                                     \mbox{5.187} \pm \mbox{0.011}\\
                                     \mbox{5.078} \pm \mbox{0.011}\\
\end{array}\right. $
& $\begin{array}{l}\mbox{3.772} \pm \mbox{0.015}\\
                            \mbox{3.649} \pm \mbox{0.015}\\
                            \mbox{3.552} \pm \mbox{0.015}\\
      \end{array} $\\
M$_{\rm bol}$\,$^{\star\star\star}$  [mag]
& $\left\{\begin{array}{l}-\mbox{8.58}\\
                                     -\mbox{8.23}\\
                                     -\mbox{7.96}\\
\end{array}\right. $
& $\begin{array}{l}-\mbox{4.69}\\
                            -\mbox{4.38}\\
                            -\mbox{4.14}\\
      \end{array} $\\
\noalign{\vskip1pt}\hline\noalign{\vskip1pt}
\multicolumn{3}{l}{$^\star$ fixed, derived from spectroscopy,}\\
\multicolumn{3}{l}{$^{\star\star}$ assumed synchronous rotation,}\\
\multicolumn{3}{l}{$^{\star\star\star}$ M$_{\rm bol,\odot}=$ 4.74~mag \citep{bess1998} was adopted.}\\
\end{tabular}
\end{minipage}

\end{table}

\begin{figure*}
\centering
\includegraphics[width=170mm]{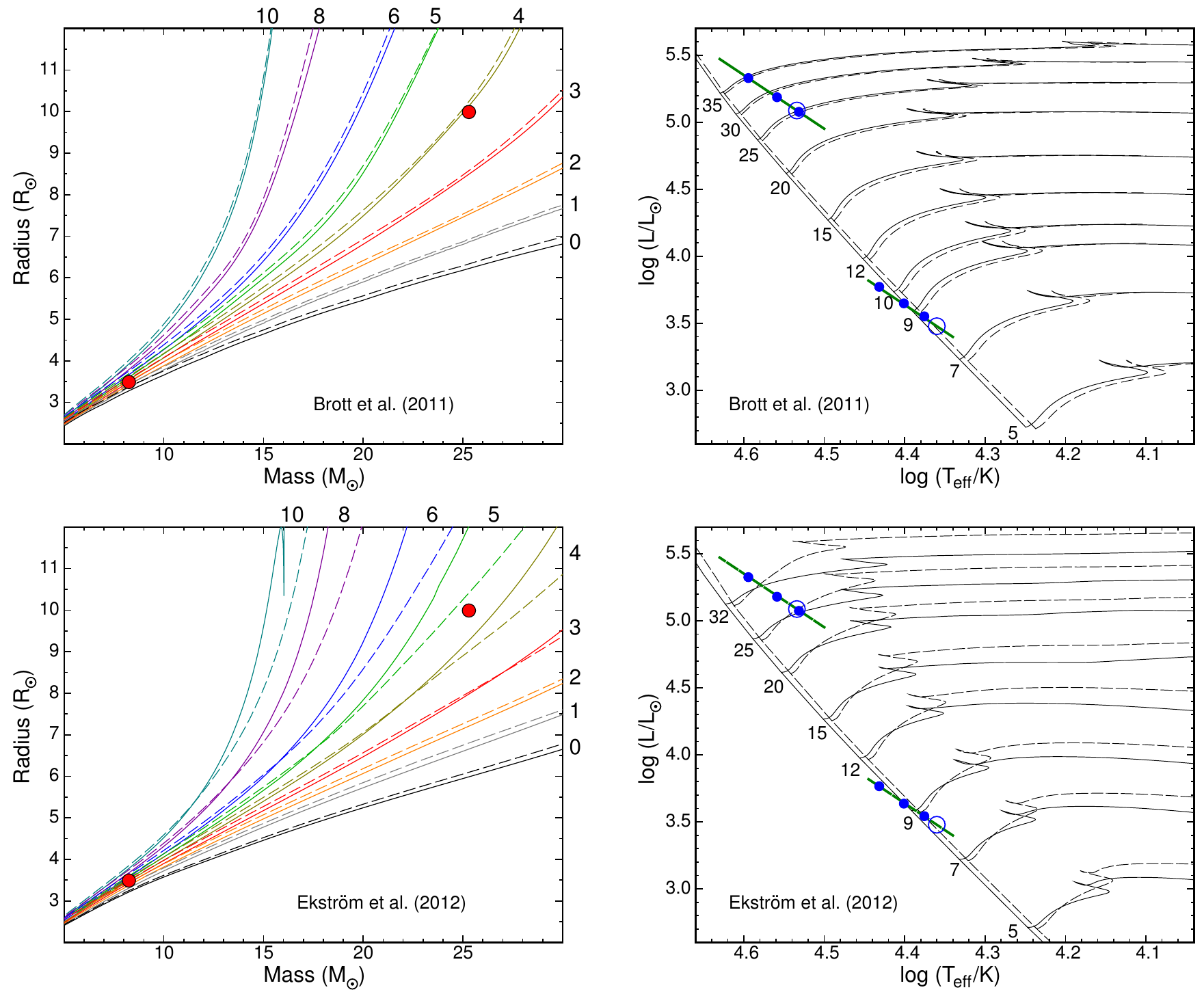}
\caption{\textit{Left:} A comparison of the dynamic masses and radii of the components of ALS\,1135 (dots) with the theoretical values taken from the stellar evolutionary models of \citet[][]{brot2011} (top) and \citet[][]{ekst2012} (bottom) for nine nine values of stellar ages. The isochrones are labeled with the age in Myr. The continuous lines denote isochrones for models without rotation, the dashed lines, isochrones with rotation. See text for more explanations. \textit{Right}: The evolutionary tracks for the same two sets of evolutionary models, \citet{brot2011} (top panel) and \citet{ekst2012} (bottom panel). 
The models are labeled with mass in $M_\odot$. As in the left-hand panels, models without rotation are shown as continuous lines, those with rotation, with dashed lines. The dots correspond to solutions for three different values of $T_{\rm eff,1}$ (see text). They are located along the lines of constant radius. Open circles denote models with a given radius and having a mass equal to the derived mass.} 
\label{mr}
\end{figure*}

The much better (multi-colour) photometry and more precise radial velocities allowed us to obtain stellar parameters of ALS\,1135 with much smaller uncertainties than those of FLN2006. In particular, the radii were derived with an accuracy better than 1\%, the masses, with an accuracy better than 3\%. Within the errors, the determined orbital period is consistent with the previous determinations (\citealt{cort2003}; FLN2006).  The main difference between the FLN2006 and our solution is the inclination, 79.2\,$\pm$\,0.2$^\circ$, about 10$^\circ$ larger than obtained by FLN2006. 
Thus, we find the eclipses to be total (see Fig.~\ref{lc1}). In consequence, the masses and radii of the components we derived are smaller than those given by FLN2006. The contribution of the secondary to the total light of the system in the visual amounts to about 6\%. The low value of the mass ratio, $M_2/M_1=$ 0.326~$\pm$~0.003, is confirmed. Assuming synchronous rotation we get $V\sin i=$\,180.5 $\pm$ 2.3~km\,s$^{-1}$ for the primary component, in very good agreement with the spectroscopic determination (see Sect.~\ref{rvap}). 

\subsection{\label{age}The age of ALS\,1135}
In their paper summarising the masses of O-type stars, \citet{wevi2010} concluded that the present-day dynamical masses of these stars are well reproduced by evolutionary models. Owing to the fact that massive stars during their main-sequence evolution considerably change their radii, there is an opportunity of using these stars (strictly speaking their radii) for determination of the ages of the stellar systems they belong to. With this purpose in mind, the masses and radii of the components of ALS\,1135 we derived were compared with those from stellar evolutionary models of \citet{brot2011} and \citet{ekst2012} (see Fig.~\ref{mr}, left-hand panels). 
In both sets of evolutionary models the rotation effects were included. One of the main differences between the models is core overshooting with $H_p$, pressure scale height, equal to 0.335 in the models of \citet{brot2011} and 0.1 in the models of \citet{ekst2012}. The much larger range of core overshooting in the models of \citet{brot2011} results in a much wider main sequence (see Fig.~\ref{mr}, right-hand panels).

The left panels in Fig.~\ref{mr} show the mass--radius relations for both sets of evolutionary models and nine values of age ranging from the zero-age main sequence up to 10~Myr. The mass-radius isochrones are plotted for models with and without rotation. In accordance with the measured $V\sin i$, we have selected models with rotation having initial rotational velocity of about 200\,km\,s$^{-1}$ for the set of \citet{brot2011} models and those with the initial rotational velocity equal to 0.4 times the critical velocity for the other set.

As can be seen from Fig.~\ref{mr}, for stars with masses larger than $\sim$15\,$M_\odot$ the radius becomes a very good indicator of the age of a star. In the ALS\,1135 system, only the primary is suitable for an age determination by this method. The ages derived from the location of this star in the mass--radius diagrams amount to 3.9~Myr for \citet{brot2011} models both with and without rotation, 4.3~Myr for \citet{ekst2012} models without rotation and 4.7~Myr for the same set of models with rotation. 
The ages obtained in a similar way for the older sets of models, \citet{clar2004} and \citet{scha1992}, resulted in the age of 4.2~Myr. The uncertainties of mass and radius (Table \ref{talspar}) are negligible in the present context. We therefore conclude that the age of the ALS\,1135 system is equal to 4.3 $\pm$ 0.5~Myr, where the adopted uncertainty of 0.5~Myr reflects only the dependability on the models. Although the result is model-dependent, it is clear that the method provides the age which is much more precise than that obtained from isochrone fitting. 

Since $T_{\rm eff,1}$ was derived from the observed spectrum, we should check the consistency of the solution obtained in Sect.~\ref{lcwd} with the evolutionary models we used to derive the age of ALS\,1135. This is done in the right-hand panels of Fig.~\ref{mr} too. The solutions for the three assumed values of $T_{\rm eff,1}$ are located, as expected, along the constant radius line. The same is true for the secondary. The models that are consistent with the derived masses are shown as open circles. It can be seen that the consistent model for the primary is located close to the solution for $T_{\rm eff,1}=$ 34\,000~K, while for the secondary it is slightly less luminous than all three solutions. Nevertheless, the consistency remains reasonable. 

\section{\label{svar}Variable stars in the observed field}
\begin{figure}
\centering
\includegraphics[width=85mm]{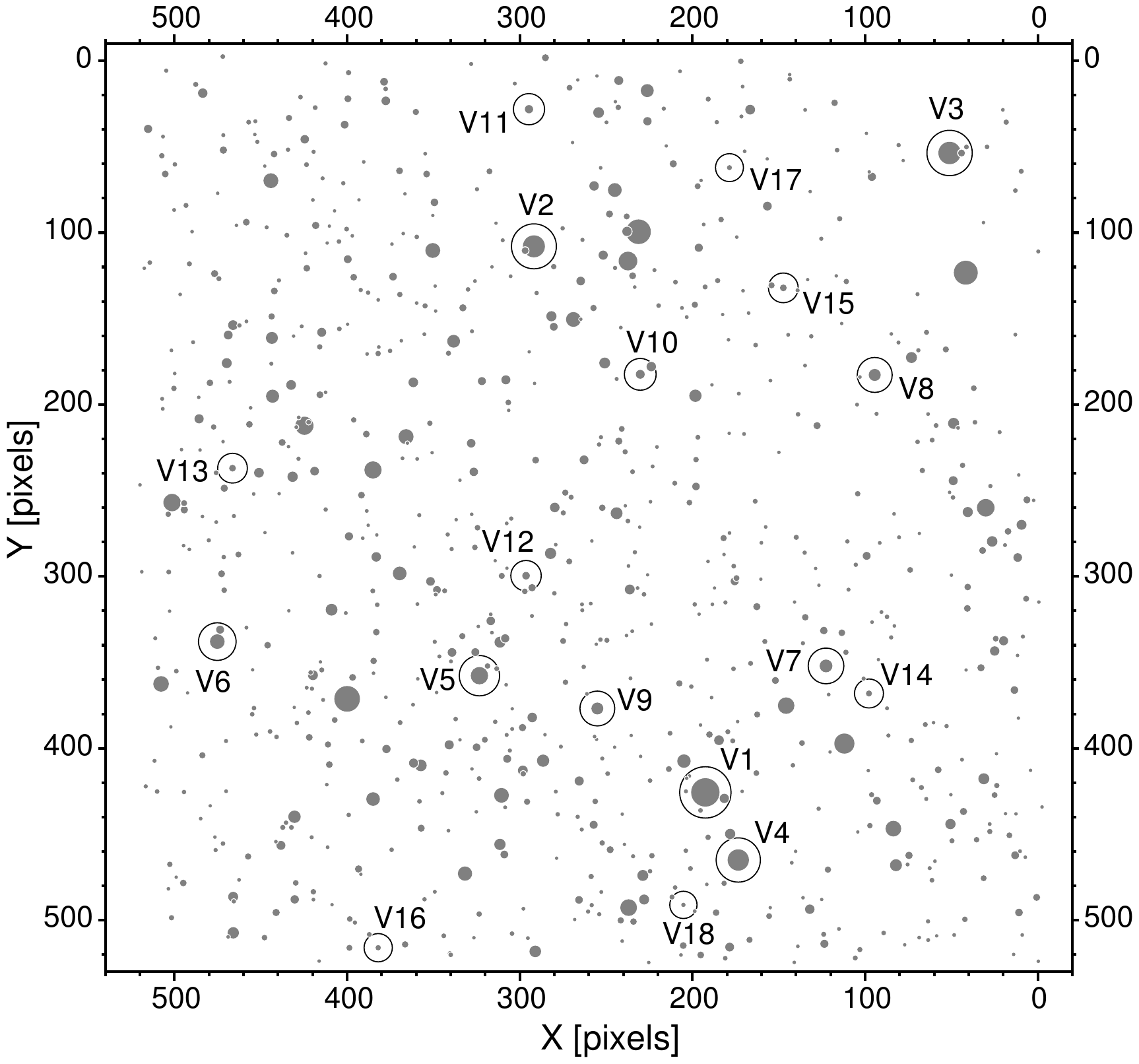}
\caption{Schematic view of the observed field with the positions of all 813 stars detected in the $I_{\rm C}$ band. Variable stars are encircled and labeled. North is up, east to the left.} 
\label{xy}
\end{figure}

As mentioned above, the photometric observations of the ALS\,1135 field were taken with two telescopes covering almost the same area of the sky. The multisite observations allow to reduce aliasing in the Fourier amplitude spectra and lower detection level. For this reason, we combined data from the SAAO and Perth observatories. Using profile photometry obtained with the DAOPHOT package, differential magnitudes of all detected stars were computed. In the search for variability, the $V$ and $I_{\rm C}$-filter magnitudes were analysed by means of a Fourier periodogram calculated in the range between 0 and 100 d$^{-1}$. Then, the light curves, the Fourier amplitude spectra and phase diagrams were inspected by eye. 
Among 813 stars detected in the $I_{\rm C}$ band, 18 were found to be variable. Only ALS 1135 (Sect.~3) was known to be variable prior to our study. The variable stars are listed in Table~\ref{tvar} and labeled in Fig.~\ref{xy}. Since the WEBDA\footnote{The WEBDA database is available at http://www.univie.ac.at/webda/.} database does not include all stars in our field, we number the variable stars from V1 to V18.

\begin{table}
 \centering
 \begin{minipage}{85mm}
  \caption{\label{tvar}A list of variable stars in the observed field. Bochum 7 is abbreviated to Bo7. The numbers in parentheses denote the r.m.s.~errors of the preceding quantities with the leading zeroes omitted. Uncertain membership is followed by a colon.}
  \begin{tabular}{@{}crlcc}
  \hline
Star  & \multicolumn{1}{c}{$V$}   &   \multicolumn{1}{c}{Period}& Membership & Type of \\
 & \multicolumn{1}{c}{[mag]} &  \multicolumn{1}{c}{[d]} & &  variability \\
\noalign{\vskip1pt}\hline\noalign{\vskip1pt}
V2*  & 11.93 &  1.133021(24)  & Vel OB1 & SPB \\
V5*  & 13.80 &  0.0420225(20)  & Vel OB1: & $\delta$ Sct \\
V6*  & 14.61 &  0.0800276(25)  & Vel OB1: & $\delta$ Sct \\
V7  & 15.40 &  0.088809(10)   & Vel OB1: & $\delta$ Sct \\
V10*  & 16.89 &  0.075080(5)  & Bo7:  PMS? & $\delta$ Sct \\
V12 & 17.43 &  0.097308(15)   & Bo7:  PMS? & $\delta$ Sct \\
V14* & 18.36 &  0.061382(11)  & field: & $\delta$ Sct \\
\noalign{\vskip1pt}\hline\noalign{\vskip1pt}
V1  & 10.90 &  2.753189(14) & Bo7 & EA (ALS1135) \\
V4  & 12.14 &  0.8654776(14) & Bo7 & EW \\
V11 & 17.11 & 4.5195(6) & unknown & EA\\
V13 & 17.80 & 0.2862813(11) & unknown & EW (W UMa)\\
V16 & 18.73 & 0.37748(5)  & unknown & EW/Ell \\
V17 & 19.54 & 0.2148565(3) & field & EW (W UMa) \\
V18 & 19.55 & 0.308876(6)  & unknown & EW/Ell \\
\noalign{\vskip1pt}\hline\noalign{\vskip1pt}
V3 & 12.10 & $\sim$10 & field & unknown \\
V8  & 15.84 & 0.57228(21)& unknown & unknown \\
V9  & 15.86 & 0.93725(26)& unknown & unknown \\
V15 & 18.72 & $\sim$10 & field & unknown \\
\noalign{\vskip1pt}\hline\noalign{\vskip1pt}
\multicolumn{5}{l}{* more than one periodicity detected, see Table \ref{tampl}.}
\end{tabular}
\end{minipage}
\end{table}

\subsection{Pulsating stars}
Out of 18 variables detected in the observed field, seven appear to be pulsating stars. One of them, V2, is probably an SPB-type star, the remaining six are $\delta$~Scuti stars. The parameters of the sinusoidal terms (frequencies, $f_i$, and amplitudes, $A_i$) for these stars, were derived by fitting the formula
\begin{equation}
{\label{eq}}
\langle m \rangle + \sum_{i=1}^{n} A_i \sin (\mbox{2} \pi f_i (t-T_0) + \phi _i)
\end{equation}
to the differential magnitudes. In Eq.~(\ref{eq}) $\langle m \rangle$ is the mean differential magnitude, $n$, the number of fitted terms, $\phi _i$, the phases, $t$, the time elapsed from the initial epoch $T_0=$ HJD\,2454400.  The parameters of the fit are listed in Table\,\ref{tampl}. Instead of $\phi _i$, we provide the time of maximum light, $T_{\rm max}$.

\begin{figure}
\centering
\includegraphics[width=7cm]{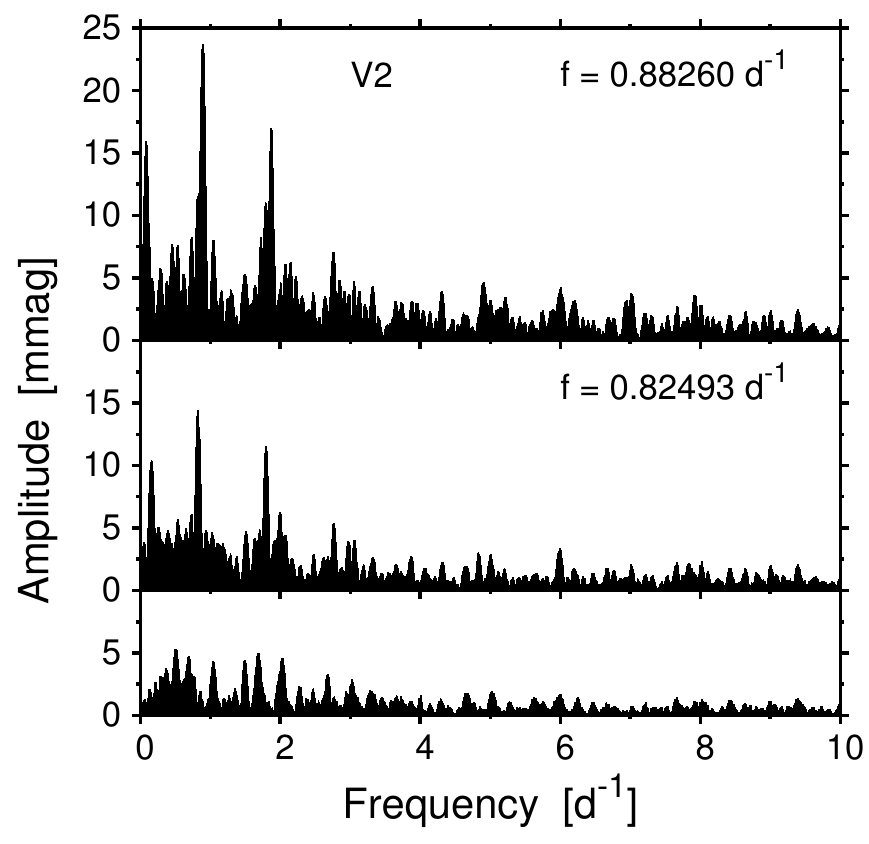}
\caption{{\it Top:} Fourier frequency spectrum of the combined SAAO and Perth $V$-filter data of the SPB star V2. {\it Middle:} after prewhitening with frequency $f_1=$ 0.88260 d$^{-1}$. {\it Bottom:} after prewhitening with both frequencies, $f_1$ and $f_2=$ 0.82493 d$^{-1}$.} 
\label{trf6}
\end{figure}

The brightest pulsator found in our data is star V2. It has no available spectral type. Fourier periodogram of the $V$-filter data for this star (Fig.\,\ref{trf6}) reveals two terms with close frequencies, $f_1=$ 0.88260 and $f_2=$ 0.82493~d$^{-1}$. The residuals indicate a possible presence of the further terms with low amplitudes and frequencies below 3~d$^{-1}$. The frequencies and the position of V2 in the colour-colour (Fig.\,\ref{color_color}) and the colour-magnitude diagrams (Fig.\,\ref{cmd}) indicate that the star is an SPB variable and a member of Vel\,OB1.

\begin{figure}
\centering
\includegraphics[width=84mm]{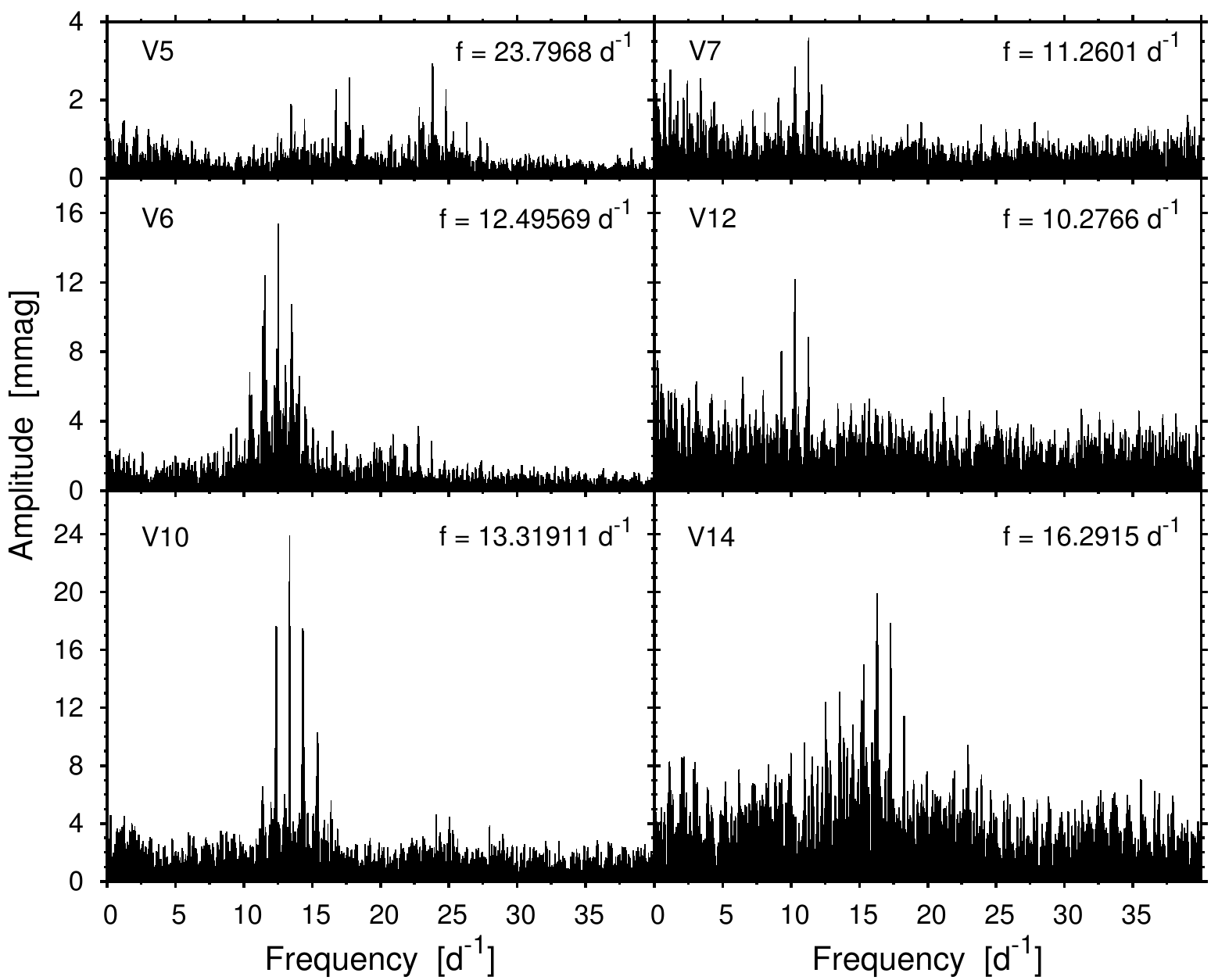}
\caption{Fourier frequency spectra of the combined SAAO and Perth $V$-filter data of six $\delta$ Scuti stars: V5, V6, V7, V10, V12 and V14.}
\label{trf14}
\end{figure}

The other six stars (V5, V6, V7, V10, V12, and V14) have periods between 0.04 and 0.18~d. Fourier periodograms of $V$-filter data of these stars are shown in Fig.~\ref{trf14}. Prewhitening original data with the strongest mode revealed more terms in four stars, V5, V6, V10, and V14. For none of these stars spectral types are available but the periods of variability, multiperiodicity of some of them and their position in the colour-magnitude diagram (Fig.\,\ref{cmd}) indicate that these stars are $\delta$ Scuti-type variables. In addition, in all these stars the amplitude decreases with increasing wavelength (Table \ref{tampl}), which is a typical property of $\delta$\,Scuti stars in the visual \citep[see, e.g.,][]{wats1988}. As discussed in Sect.\,\ref{scmd}, stars V5, V6 and V7 are likely members of the Vel OB1 association, whereas V10 and V12 probably belong to Bochum\,7. V14 seems to be a field star.

\begin{table*}
 \centering
 \begin{minipage}{140mm}
  \caption{\label{tampl}Parameters of sine-curve fits to the $B$, $V$ and $I_{\rm C}$ differential magnitudes of the pulsating stars detected in the observed field. The numbers in parentheses denote the r.m.s.~errors of the preceding quantities with the leading zeroes omitted. The $\sigma_{\rm res}$ is the residual standard deviation, $N_{\rm obs}$ stands for the number of observations, $S/N$ is the signal-to-noise ratio.}
  \begin{tabular}{@{}crccrcrr}
  \hline
Star & \multicolumn{1}{c}{$f_i$} & Filter & $N_{\rm obs}$& \multicolumn{1}{c}{$A_i$} & \multicolumn{1}{c}{$T_{\rm max} - T_0$} & \multicolumn{1}{c}{$S/N$} &\multicolumn{1}{c}{$\sigma_{\rm res}$} \\
& \multicolumn{1}{c}{[d$^{-1}$]}& & & \multicolumn{1}{c}{[mmag]} & \multicolumn{1}{c}{[d]} & &\multicolumn{1}{c}{[mmag]}  \\\hline\noalign{\vskip1pt} 
V2  & 0.88260(19)  & $B$ & 1581 & 31.8(6)  & 95.4740(32) & 44.9 & 14.8 \\
    &              & $V$ & 2456 & 26.5(3)  & 96.6303(21) & 72.1 &  9.4 \\
    &              & $I$ & 2670 & 22.3(4)  & 96.6128(28) & 72.1 & 12.0 \\
    & 0.82493(32)  & $B$ &      & 22.0(6)  & 95.8864(53) & 31.0 &      \\
    &              & $V$ &      & 17.4(4)  & 95.9055(33) & 47.4 &      \\
    &              & $I$ &      & 13.7(4)  & 95.8980(59) & 26.6 &      \\\noalign{\vskip1pt}\hline\noalign{\vskip1pt}
V5  & 23.7968(11)  & $B$ & 1575 &  3.9(3)  & 95.5939(05) &  9.5 &  7.6 \\
    &              & $V$ & 2457 &  2.9(2)  & 96.0962(04) & 10.0 &  6.6 \\
    &              & $I$ & 2656 &  1.7(2)  & 96.1812(07) &  6.1 &  6.7 \\
    & 17.7011(11)  & $B$ &      &  2.8(3)  & 95.6216(09) &  6.8 &      \\
    &              & $V$ &      &  2.5(2)  & 96.0739(07) &  8.9 &      \\
    &              & $I$ &      &  1.4(2)  & 96.1832(11) &  5.0 &      \\
    & 13.4460(21)  & $B$ &      &  1.6(2)  & 95.5796(29) &  4.0 &      \\
    &              & $V$ &      &  1.9(2)  & 96.1020(12) &  6.6 &      \\
    &              & $I$ &      &  1.3(2)  & 96.1817(17) &  4.7 &      \\
    & 26.2978(23)  & $B$ &      &  1.8(3)  & 95.5951(09) &  4.5 &      \\
    &              & $V$ &      &  1.4(2)  & 96.0878(08) &  4.8 &      \\
    & 23.0559(39)  & $V$ &      &  1.3(2)  & 96.0803(10) &  4.4 &      \\
    & 17.5605(25)  & $B$ &      &  1.8(3)  & 95.6209(14) &  4.5 &      \\
    &              & $I$ &      &  1.2(2)  & 96.1885(14) &  4.4 &      \\\noalign{\vskip1pt}\hline\noalign{\vskip1pt}
V6  & 12.49569(39) & $B$ & 1558 & 23.3(6)  & 95.6425(03) & 28.0 & 17.1 \\
    &              & $V$ & 2446 & 17.8(4)  & 95.8820(03) & 33.4 & 12.6 \\
    &              & $I$ & 2671 & 10.5(4)  & 96.1227(04) & 36.0 & 12.5 \\
    & 11.40806(62) & $B$ &      & 15.9(6)  & 95.5887(05) & 19.0 &      \\
    &              & $V$ &      & 12.0(0)  & 95.8508(04) & 22.5 &      \\
    &              & $I$ &      &  7.3(3)  & 96.1140(07) & 15.5 &      \\
    & 13.03861(94) & $B$ &      &  9.4(6)  & 95.5950(08) & 11.3 &      \\
    &              & $V$ &      &  7.3(4)  & 95.9029(06) & 13.7 &      \\ 
    &              & $I$ &      &  4.9(3)  & 96.1333(09) & 10.4 &      \\
    & 22.7290(19)  & $B$ &      &  4.6(6)  & 95.6104(10) &  5.5 &      \\ 
    &              & $V$ &      &  3.2(4)  & 95.8755(08) &  6.0 &      \\ 
    &              & $I$ &      &  1.9(3)  & 96.1400(12) &  4.1 &      \\
    & 20.8858(19)  & $B$ &      &  4.7(6)  & 95.6403(10) &  5.7 &      \\ 
    &              & $V$ &      &  3.3(4)  & 95.8782(08) &  6.1 &      \\ 
    &              & $I$ &      &  2.6(3)  & 96.1181(10) &  5.5 &      \\
    & 15.7539(34)  & $B$ &      &  3.4(6)  & 95.6084(17) &  4.1 &      \\ 
    &              & $V$ &      &  2.4(4)  & 95.8591(15) &  4.6 &      \\\noalign{\vskip1pt}\hline\noalign{\vskip1pt} 
V7  & 11.2601(13)  & $B$ & 1567 &  5.5(6)  & 95.4028(15) &  6.3 & 17.6 \\
    &              & $V$ & 2471 &  3.6(4)  & 96.1147(16) &  7.1 & 13.4 \\
    &              & $I$ & 2690 &  2.2(4)  & 96.1828(23) &  4.1 & 13.9 \\\noalign{\vskip1pt}\hline\noalign{\vskip1pt}
V10 & 13.31911(83) & $B$ & 1535 & 31.1(21) & 95.8187(08) & 10.9 & 59.0 \\ 
    &              & $V$ & 2458 & 21.9(11) & 96.1191(05) & 17.4 & 33.0 \\  
    &              & $I$ & 2672 & 13.6(07) & 96.1208(06) & 12.7 & 25.0 \\  
    & 15.3988(31)  & $V$ &      &  8.2(09) & 96.1129(12) &  6.5 &      \\  
    & 12.3732(54)  & $V$ &      &  5.0(11) & 96.1695(32) &  4.0 &      \\\noalign{\vskip1pt}\hline\noalign{\vskip1pt}  
V12 & 10.2766(15)  & $B$ & 1524 & 19.0(33) & 95.8268(27) &  4.3 & 92.1 \\
    &              & $V$ & 2450 & 12.5(14) & 96.1177(17) &  6.3 & 48.8 \\
    &              & $I$ & 2672 &  6.3(09) & 97.3831(21) &  5.4 & 31.2 \\\noalign{\vskip1pt}\hline\noalign{\vskip1pt}
V14 & 16.2915(28)  & $B$ &  684 & 31.1(52) & 97.8637(17) &  4.9 &107.1 \\ 
    &              & $V$ & 1682 & 20.9(20) & 97.3739(10) &  7.7 & 56.5 \\ 
    &              & $I$ & 1892 & 15.3(13) & 97.2499(08) &  9.1 & 38.9 \\ 
    & 13.5197(42)  & $V$ &      & 16.8(20) & 97.3548(14) &  6.2 &      \\ 
    &              & $I$ &      & 10.0(13) & 97.2779(16) &  6.0 &      \\ 
    & 13.8055(64)  & $V$ &      & 11.2(20) & 97.4214(14) &  4.4 &      \\ 
    &              & $I$ &      & 10.0(13) & 97.2761(28) &  4.0 &      \\\noalign{\vskip1pt}\hline\noalign{\vskip1pt} 
\end{tabular}
\end{minipage}
\end{table*}  

\subsection{\label{secl}Eclipsing systems}
The six eclipsing binaries we found are V4, V11, V13, V16, V17, and V18. Their periods, times of the primary minimum and the depths of the eclipses are given in Table \ref{tmin}.

Star V4 (GSC\,08151-01072) is an EW-type eclipsing system with an orbital period equal to 0.8654776\,d, exactly twice the secondary period found in the ASAS data of ALS\,1135 by \citet{pimi2007}. This is the star that contaminates the variability of ALS\,1135 because it is only 27$^{\prime\prime}$ apart (Fig.\,\ref{xy}). According to \citet{cort2007}, the spectral type of this star is B1-1.5\,V and the true distance modulus is equal to 12.9 mag. This implies that the star can be a member of Bochum\,7. The $(B-V)$ and $(U-B)$ colour indices and the $V$ magnitude determined by these authors are equal to 0.40 mag, $-$0.47 mag and 12.3 mag, respectively. The out-of-eclipse $UBVI_{\rm C}$ photometry of this star obtained from our observations is given in Table\,\ref{tubvi}.

\begin{figure}
\centering
\includegraphics[width=84mm]{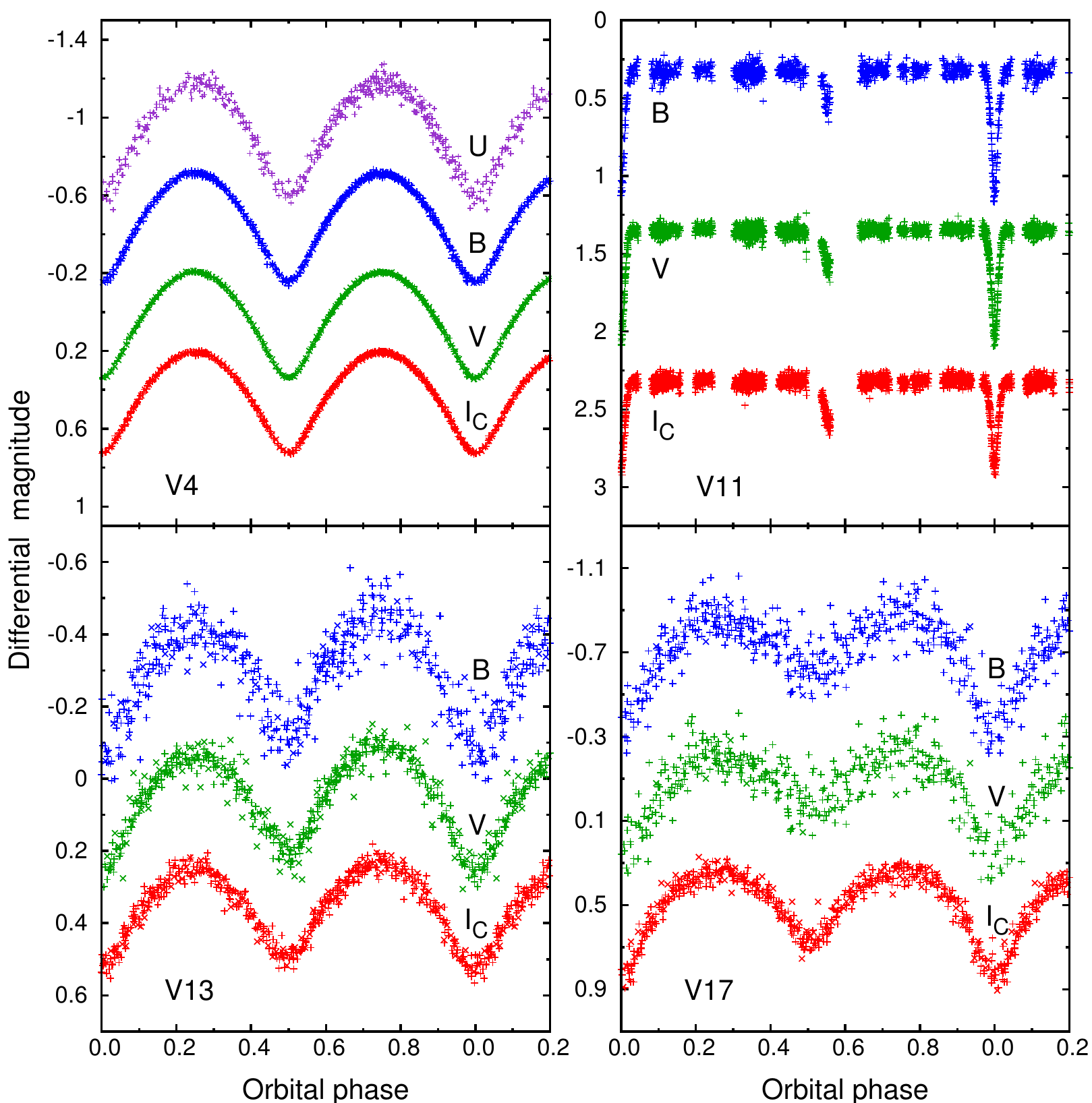}
\caption{Phase diagram of the $UBVI_{\rm C}$ observations of the eclipsing system V4 and $BVI_{\rm C}$ observations of eclipsing systems V11, V13 and V17. The SAAO data are shown as plus signs, and the Perth data as crosses. Offsets were applied to separate light curves in different bands.} 
\label{lc7}
\end{figure}

We made an attempt at modelling the light curve of V4 by means of the W-D program. The analysis revealed that both stars fill their Roche lobes. Unfortunately, this is the only firm conclusion that comes from the fit; the parameters of this system cannot be derived unambiguously as there are many equally good solutions with different mass ratios. The $U$, $B$, $V$ and $I_{\rm C}$ light curves of V4 are shown in Fig.\,\ref{lc7}. For the SAAO data, the standard deviations of the solutions are equal to 0.037, 0.011, 0.008 and 0.010~mag in $U$, $B$, $V$ and $I_{\rm C}$, respectively, while for the Perth data the respective numbers are equal to 0.014 ($B$), 0.009 ($V$) and 0.012~mag ($I_{\rm C}$).

\begin{table}
 \centering
 \begin{minipage}{85mm}
  \caption{\label{tmin}Parameters of eclipsing binary stars found in the observed field. The depths of eclipses are given for the $V$-filter observations.
  The numbers in parentheses denote the r.m.s.~errors of the preceding quantities with the leading zeroes omitted.}
  \begin{tabular}{@{}clrcc}
  \hline
Star  & \multicolumn{1}{c}{Period} & \multicolumn{1}{c}{$T_{\rm min} \mbox{I}-T_0$} & \multicolumn{2}{c}{Eclipse depth [mag]} \\
 & \multicolumn{1}{c}{[d]} & \multicolumn{1}{c}{[d]} & primary & secondary  \\
\noalign{\vskip1pt}\hline\noalign{\vskip1pt}
V4  & 0.8654776(14) & 89.7273(02) & 0.54 & 0.53 \\
V11 & 4.5195(6) & 102.4300(11) & 0.67 & $\geq$ 0.24 \\
V13 & 0.2862813(11) & 88.8412(03) & 0.34 & 0.29 \\
V16 & 0.37748(5)    & 98.4788(13) & 0.14 & 0.14  \\
V17 & 0.2148565(3)  & 88.9822(01) & 0.49 & 0.33 \\
V18 & 0.30876(6)    & 97.6034(21) & 0.12 & 0.12  \\
\noalign{\vskip1pt}\hline
\end{tabular}
\end{minipage}
\end{table}

\begin{figure}
\centering
\includegraphics[width=84mm]{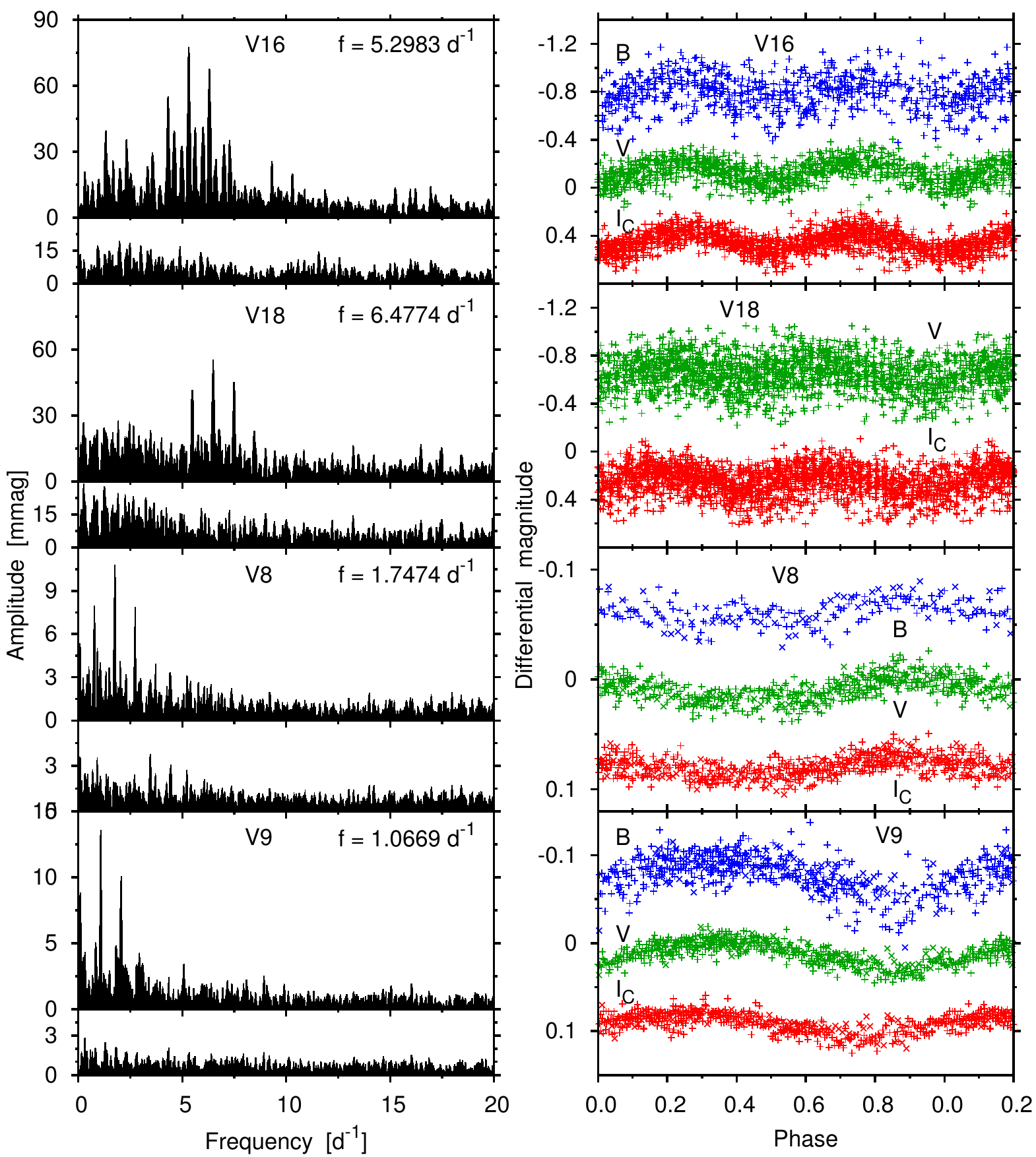}
\caption{{\it Left:} Fourier frequency spectra of the SAAO $V$-filter data of variable stars V16, V18, V8, and V9. Two frequency spectra are shown for each star: the upper for the original data and the lower after prewhitening with detected frequencies. {\it Right:} The phase diagrams in $B$, $V$ and $I_{\rm C}$. For V16 and V18 we adopted periods equal to $\mbox{2}/f$, where $f$ is the frequency. Offsets were applied in order to separate light curves in different bands.} 
\label{trflc311}
\end{figure}

The next three eclipsing systems found in our field (V11, V13 and V17) have no available spectral types. Their periods and parameters of the eclipses are given in Table \ref{tmin}. The light curves of V13 and V17 in $B$, $V$ and $I_{\rm C}$ filters are shown in Fig.~\ref{lc7}. The out-of-eclipse $V$ magnitudes and the $(B-V)$ and $(V-I_{\rm C})$ colour indices of these stars obtained from our observations are shown in Table\,\ref{tubvi}. The orbital periods and the light curves of V13 and V17 are typical for W\,UMa-type stars. On the other hand, V11 has the longest orbital period of the eclipsing stars we discovered. As can be seen in Fig.~\ref{lc7} this is a well detached EA-type system with an eccentric orbit (the secondary eclipse is not exactly in phase 0.5).

In the power spectra of the remaining two variables, V16 and V18, we have detected single periodicities with frequencies $f$ equal to 5.2983 and 6.4774\,d$^{-1}$, respectively (Fig.~\ref{trflc311}). Although these frequencies may indicate $\delta$\,Scuti-type variability, the stars do not show the decrease of the amplitude with increasing wavelength, characteristic of $\delta$\,Scuti stars. Therefore, we suppose that they are EW-type binary systems with small orbital inclinations and orbital periods equal to 2$/f$. Another argument in favour of this interpretation is that their amplitudes are quite large as for $\delta$\,Scuti stars.

\subsection{Other variables}
\begin{figure}
\centering
\includegraphics[width=84mm]{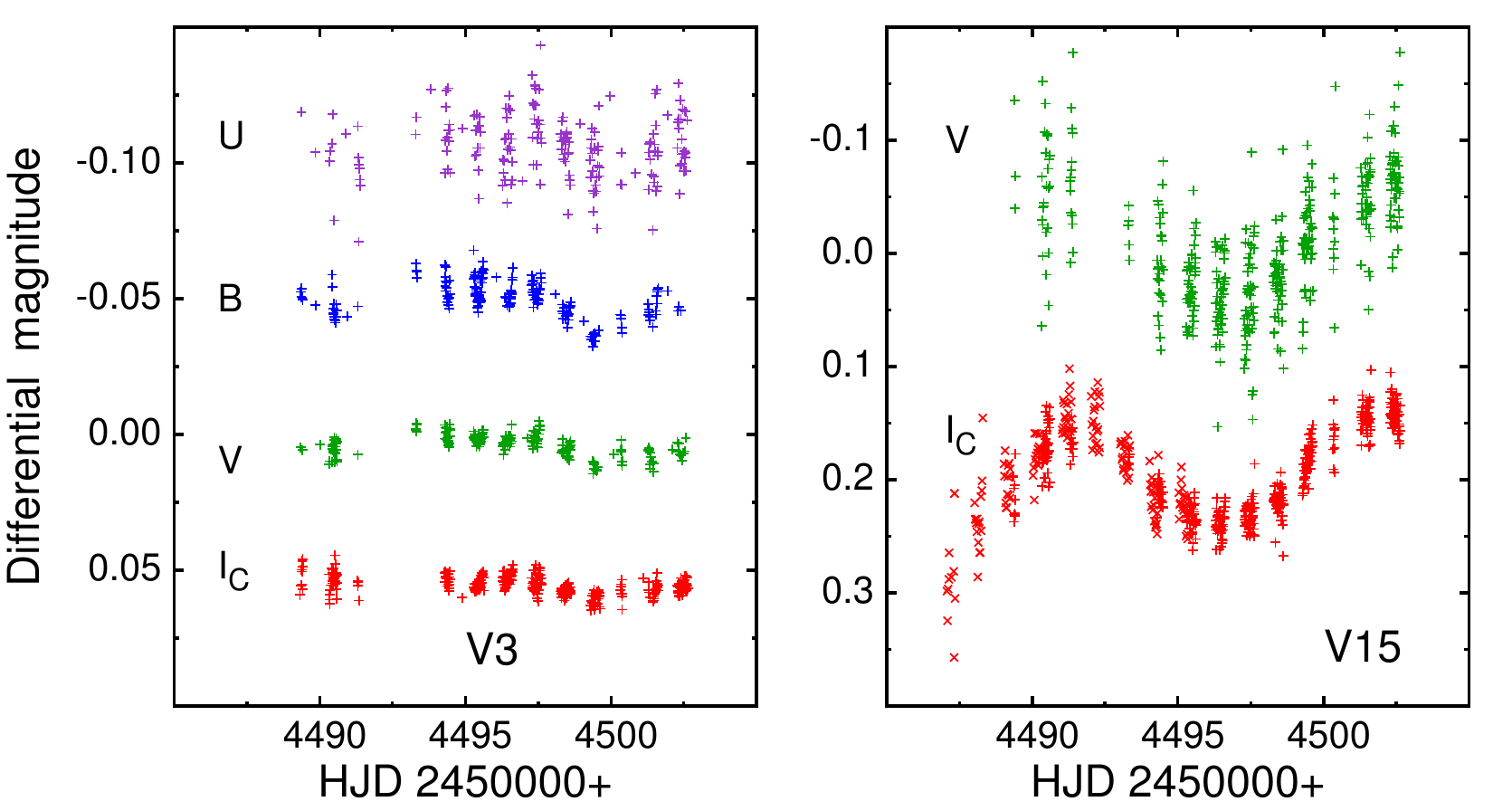}
\caption{The light curves of variable stars V3 (left) and V15 (right) of unknown type. The SAAO data are shown as plus signs, the Perth data, as crosses.} 
\label{lc127}
\end{figure}

In addition, we detected variability in four other stars: V3, V8, V9 and V15. Our data do not allow us to classify unambiguously their variability. The highest peak in the periodogram of V8 is at $f=$ 1.7474~d$^{-1}$ (Fig.~\ref{trflc311}). Similarly, the variation of star V9 can be described by a single frequency equal to 1.0669~d$^{-1}$ (Fig.~\ref{trflc311}). The third star, V3, is a field star showing quasi-periodic variability with a period of about 10\,d (Fig.\,\ref{lc127}). The fourth star, V15, is very red (Fig.\,\ref{cmd}) showing variations on a time-scale of the order of 10\,d. Its light curves in $V$ and $I_{\rm C}$ are shown in Fig.\,\ref{lc127}. 

\section{\label{subvi}$UBVI_{\rm C}$ photometry}
As mentioned in the Introduction, ALS\,1135 is situated in a region rich in young OB associations. For instance, \citet{reed1988,reed1990} confirmed the presence of Vel\,OB1, OB2 and OB3 (Bochum 7) at distances of 2.1, 0.65 and 5.3\,kpc and pointed out the presence of fourth association, Vel\,OB4, at a distance of 1.0 kpc. As was suggested by \citet{kal2000}, the situation can be more intricate. These authors analysed the distribution of bright OB stars in the Vela region and found even more clumps of stars along the same line of sight. According to \citet{sung1999} Vel\,OB1 is situated at a distance of 1.8\,kpc. The same authors obtained a distance of 4.8~kpc for Vel\,OB3. The most recent determinations of the distance to Vel\,OB3 was given by \citet{cort2003}, who derived a distance of 5.0~kpc from the analysis of radial velocities of several OB-type members of this association.

We observed a small part of this interesting region and were able to identify stars from only two associations, Vel\,OB1 and Vel \,OB3 (Bochum\,7). Stars from the nearest association, Vel\,OB2, are scattered in a large field, and none falls in the observed field of view.

\begin{table*}
 \centering
 \begin{minipage}{120mm}
  \caption{\label{tubvi}$UBVI_{\rm C}$ photometry and coordinates of variable stars.}
  \begin{tabular}{@{}rccccccr}
  \hline
Star & USNO-B1.0 & R.Asc.\,(2000.0) & Dec.\,(2000.0) & $V$  & $(V-I_{\rm C})$& $(B-V)$ & $(U-B)$ \\
\noalign{\vskip1pt}\hline\noalign{\vskip1pt}
V1 & 0438-0140625 & 8$^{\rm h}$43$^{\rm m}$49.83$^{\rm s}$ & $-$46$^{\rm o}$07$^{\prime}$08.8$^{\prime\prime}$ & 10.898 & 0.623 & 0.358 & $-$0.557 \\ 
V2 & 0439-0140368 & 8$^{\rm h}$43$^{\rm m}$54.74$^{\rm s}$ & $-$46$^{\rm o}$03$^{\prime}$49.9$^{\prime\prime}$ & 11.927 & 0.387 & 0.247 & $-$0.178 \\ 
V3 & 0439-0140136 & 8$^{\rm h}$43$^{\rm m}$40.24$^{\rm s}$ & $-$46$^{\rm o}$03$^{\prime}$23.6$^{\prime\prime}$ & 12.103 & 0.787 & 0.642 & 0.040 \\ 
V4 & 0438-0140604 & 8$^{\rm h}$43$^{\rm m}$48.80$^{\rm s}$ & $-$46$^{\rm o}$07$^{\prime}$33.6$^{\prime\prime}$ & 12.139 & 0.676 & 0.416 & $-$0.400 \\ 
V5 & 0438-0140769 & 8$^{\rm h}$43$^{\rm m}$57.39$^{\rm s}$ & $-$46$^{\rm o}$06$^{\prime}$22.9$^{\prime\prime}$ & 13.804 & 0.671 & 0.477 & 0.160 \\ 
V6 & 0438-0140922 & 8$^{\rm h}$44$^{\rm m}$06.32$^{\rm s}$ & $-$46$^{\rm o}$06$^{\prime}$05.5$^{\prime\prime}$ & 14.611 & 0.723 & 0.542 & 0.252 \\ 
V7 & 0438-0140540 & 8$^{\rm h}$43$^{\rm m}$45.47$^{\rm s}$ & $-$46$^{\rm o}$06$^{\prime}$25.7$^{\prime\prime}$ & 15.398 & 0.757 & 0.571 & 0.229 \\ 
V8 & 0439-0140186 & 8$^{\rm h}$43$^{\rm m}$43.27$^{\rm s}$ & $-$46$^{\rm o}$04$^{\prime}$42.1$^{\prime\prime}$ & 15.836 & 1.145 & 0.920 & 0.385 \\ 
V9 & 0438-0140694 & 8$^{\rm h}$43$^{\rm m}$53.40$^{\rm s}$ & $-$46$^{\rm o}$06$^{\prime}$36.8$^{\prime\prime}$ & 15.864 & 1.130 & 0.896 & 0.247 \\ 
V10 & 0439-0140303 & 8$^{\rm h}$43$^{\rm m}$51.33$^{\rm s}$ & $-$46$^{\rm o}$04$^{\prime}$37.7$^{\prime\prime}$ & 16.885 & 1.135 & 0.801 & 0.249 \\ 
V11 & 0439-0140365 & 8$^{\rm h}$43$^{\rm m}$54.66$^{\rm s}$ & $-$46$^{\rm o}$03$^{\prime}$00.6$^{\prime\prime}$ & 17.110 & 1.209 & 0.852 & 0.475\\ 
V12 & --- & 8$^{\rm h}$43$^{\rm m}$55.61$^{\rm s}$ & $-$46$^{\rm o}$05$^{\prime}$47.9$^{\prime\prime}$ & 17.427 & 1.226 & 0.920 & 0.290 \\ 
V13 & 0439-0140568 & 8$^{\rm h}$44$^{\rm m}$05.48$^{\rm s}$ & $-$46$^{\rm o}$05$^{\prime}$03.5$^{\prime\prime}$ & 17.799 & 1.231 & 1.070 & --- \\ 
V14 & 0438-0140522 & 8$^{\rm h}$43$^{\rm m}$44.05$^{\rm s}$ & $-$46$^{\rm o}$06$^{\prime}$36.4$^{\prime\prime}$ & 18.355 & 1.515 & 1.001 & --- \\ 
V15 & 0439-0140229 & 8$^{\rm h}$43$^{\rm m}$46.25$^{\rm s}$ & $-$46$^{\rm o}$04$^{\prime}$09.1$^{\prime\prime}$ & 18.720 & 2.617 & ---      & --- \\ 
V16 & 0438-0140834 & 8$^{\rm h}$44$^{\rm m}$01.36$^{\rm s}$ & $-$46$^{\rm o}$07$^{\prime}$58.7$^{\prime\prime}$ & 18.730 & 1.343 & 0.988 & --- \\ 
V17 & 0439-0140249 & 8$^{\rm h}$43$^{\rm m}$47.86$^{\rm s}$ & $-$46$^{\rm o}$03$^{\prime}$25.0$^{\prime\prime}$ & 19.535 & 2.348 & ---      & --- \\ 
V18 & 0438-0140645 & 8$^{\rm h}$43$^{\rm m}$50.76$^{\rm s}$ & $-$46$^{\rm o}$07$^{\prime}$48.7$^{\prime\prime}$ & 19.549 & 1.503 & ---      & --- \\ 
\noalign{\vskip1pt}\hline
\end{tabular}
\end{minipage}
\end{table*} 

\subsection{Transformation to the standard system}
The $UBVI_{\rm C}$ photometry of Bochum 7 and Vel OB1 was provided by \citet{sung1999}. Our field of view contains 29 stars observed by these authors. For several bright stars in our field, $(B-V)$ and $(U-B)$ colour indices are also listed by \citet{cort2003}. Using these stars as standards, the following transformation equations were derived:
\begin{equation}
\label{eq_V}
V=v+(\mbox{0.007} \pm \mbox{0.011})\times(v-i)+(\mbox{11.965} \pm \mbox{0.007}),
\end{equation}
\begin{equation}
V-I_{\rm C}=(\mbox{0.957} \pm \mbox{0.012})\times(v-i)+(\mbox{0.757} \pm \mbox{0.008}),
\end{equation}
\begin{equation}
B-V=(\mbox{0.943} \pm \mbox{0.007})\times(b-v)+(\mbox{0.430} \pm \mbox{0.005}),
\end{equation}
\begin{equation}
\label{eq_UB}
U-B=(\mbox{1.021} \pm \mbox{0.032})\times(u-b)-(\mbox{0.189} \pm \mbox{0.020}),
\end{equation}
where $u$, $b$, $v$ and $i$ denote the mean instrumental magnitudes for the SAAO data. For eclipsing stars the transformed magnitudes correspond to the phases of maximum light, for V3 and V15 they correspond to the epoch of HJD\,2454496.5. The residual standard deviations for the transformation equations were the following: 0.025, 0.027, 0.020 and 0.079~mag for Eqs.~\ref{eq_V}--\ref{eq_UB}, respectively. The coefficients for the colour terms are close either to 0 (Eq.~\ref{eq_V}) or 1 (the remaining equations). This means that the instrumental system reproduces reasonably well the standard one. 
From the above equations, we computed $V$ magnitudes and the $(V-I_{\rm C})$ colour indices for about 630 stars and the $(U-B)$ and $(B-V)$ colour indices for about 130 and 400 stars, respectively. The standard photometry of the variable stars is given in Table\,\ref{tubvi}\footnote{The full version of Table\,\ref{tubvi} is available in electronic form from the CDS.}. The equatorial coordinates in this table were calculated from average stellar positions by means of an astrometric transformation to the positions of 201 stars in the USNO-B1.0 catalog \citep{mone2003}.

\begin{figure*}
\centering
\includegraphics[width=12.7cm]{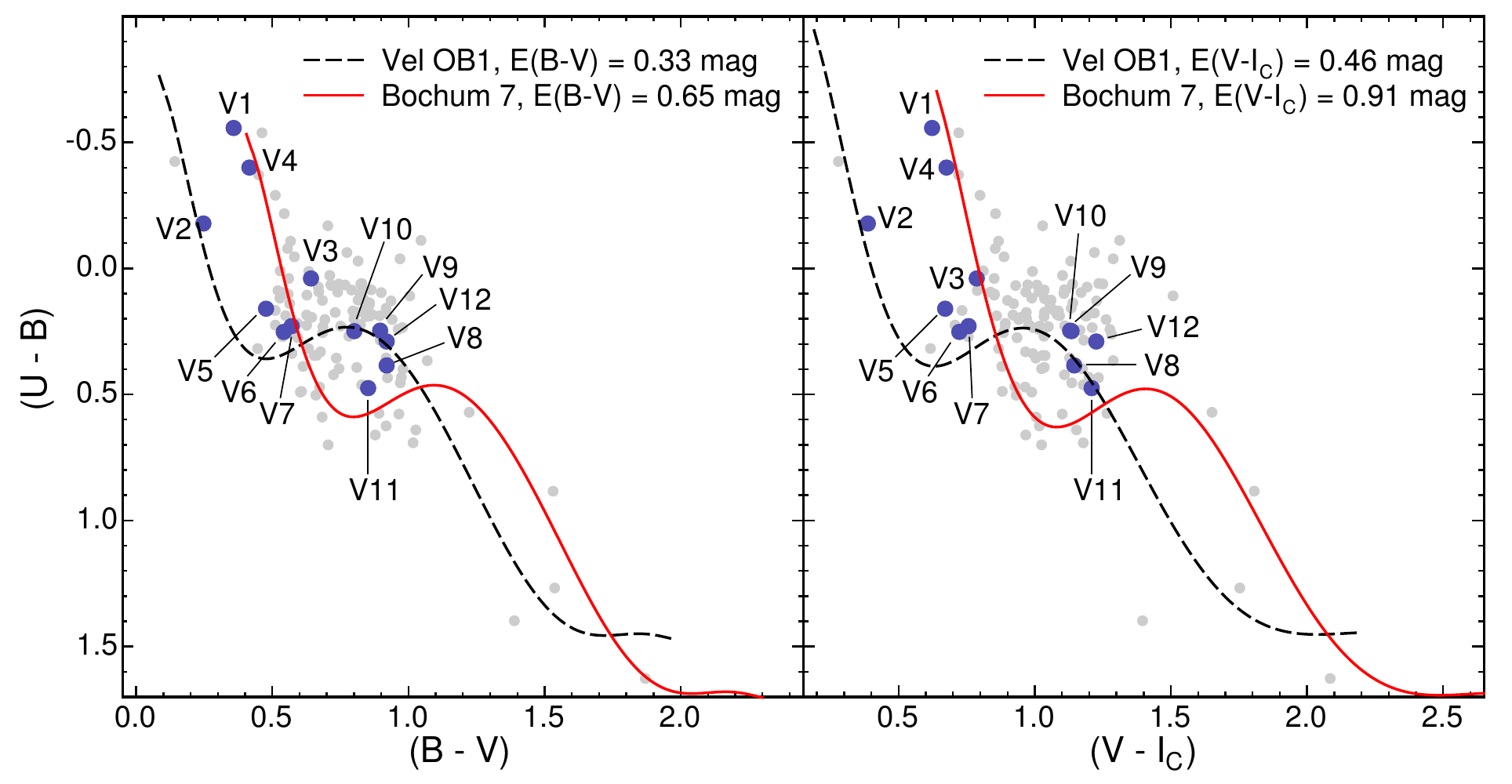}
\caption{The two-colour diagrams, $(U-B)-(B-V)$ (left) and $(U-B)-(V-I_{\rm C})$ (right) for the observed field. The variable stars are shown as large filled circles and labeled. The lines represent two-colour relations for dwarfs taken from \citet{cald1993} and shifted by the indicated mean values of colour excesses for Vel OB1 (dashed line) and Bochum 7 (solid line). See text for details.}
\label{color_color}
\end{figure*}

\subsection{\label{scmd}Colour-colour and colour-magnitude diagrams}
As mentioned above, the field we observed contains stars from at least two OB associations, Vel OB1 and Bochum 7. \citet{sung1999} and \citet{cort2007} observed much larger field than we did, 30$^\prime$ $\times$ 30$^\prime$, roughly covering the area shown in Fig.~\ref{map}. As found by \citet{sung1999}, the reddenings of these two stellar groups are considerably different and vary over the field. The $E(B-V)$ colour excess amounts to 0.20--0.45~mag for Vel OB1 and 0.75--1.05~mag for Bochum 7. From the $UBV$ observations of practically the same field made by \citet{cort2007} the following ranges of $E(B-V)$ can be estimated: 0.25--0.48~mag for Vel~OB1 and 0.58--1.0~mag for Bochum 7, in good agreement with \citet{sung1999}. It is obvious that the spread of reddening in our small field will be much smaller. 
Indeed, in the colour-colour diagrams (Fig.~\ref{color_color}) there are two groups of OB stars that can be identified as members of Vel OB1 and Bochum 7. Assuming $E(U-B)/E(B-V) =$ 0.72, we derived their mean $E(B-V)$ colour-excesses; they amount to 0.33~mag for Vel~OB1 (from two stars) and 0.65~mag for Bochum 7 (from six stars). The latter value does not agree with 0.95~mag read off the reddening map shown by \citet[][see their fig.~4, south-west corner]{sung1999}. This discrepancy can be explained by a low number of members of Bochum 7 used to calculate the reddening map. The eight certain OB members of both associations located in our field are listed in Table \ref{OBstars}.

\begin{table}
 \centering
 \begin{minipage}{85mm}
  \caption{\label{OBstars}OB stars in the observed field.}
  \begin{tabular}{@{}clcl}
  \hline
\multicolumn{1}{c}{Star} & \multicolumn{1}{c}{$V$} & Member & Remarks\\
\noalign{\vskip1pt}\hline\noalign{\vskip1pt}
ALS 1135 & 10.898  & Bochum 7 & O6.5 V((f)), V1\\ 
ALS 1137  & 11.407 & Bochum 7 & O9-9.5 V\\
CBN84348.6-460736 &  12.139  & Bochum 7 & B1-1.5 V, V3\\     
CBN84344.7-460656 & 12.925   & Bochum 7 & B1 V\\
CBN84346.7-460641 & 14.157   & Bochum 7 & B1-5 V\\
2MASS 08440123-4607049 & 15.161 &  Bochum 7 & \\
\noalign{\vskip1pt}\hline\noalign{\vskip1pt}
CPD $-$45$^\circ$2913 & 11.136  & Vel OB1 & \\
CPD $-$45$^\circ$2922 & 11.927  & Vel OB1 & V2\\
\noalign{\vskip1pt}\hline
\end{tabular}
\end{minipage}
\end{table}

In a similar way, mean $E(V-I_{\rm C})$ colour-excesses were derived using data for eight stars from Table \ref{OBstars}. We assumed $E(U-B)/E(V-I_{\rm C})=$ 0.536, a consequence of adopting $E(V-I_{\rm C})/E(B-V)=$ 1.343 derived by \citet{sung1999}. We obtained $E(V-I_{\rm C}) =$ 0.46~mag for Vel OB1 and 0.91~mag for Bochum 7 (Fig.~\ref{color_color}). Comparing colour excesses we get $E(V-I_{\rm C})/E(B-V)\approx$ 1.4 for our field. Bearing in mind the small number of stars we used in our calculation, this value can be regarded as consistent with that given by \citet{sung1999}.

\begin{figure*}
\centering
\includegraphics[width=10.5cm]{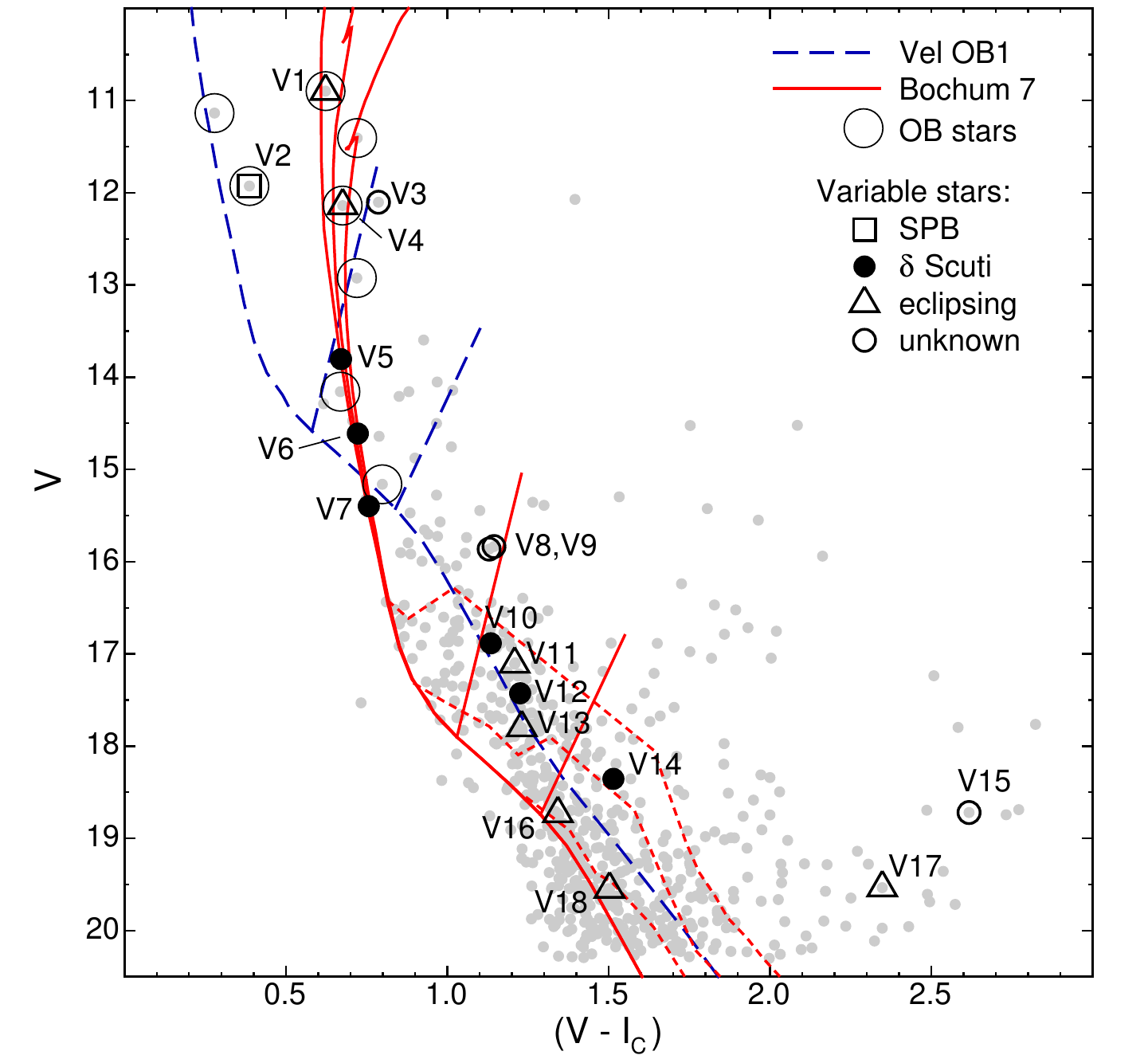}
\caption{ The $V$ vs.~$(V-I_{\rm C})$ colour-magnitude diagram for the observed field. Variable stars are shown with different symbols indicating the type of variability and labeled (see text for more information). The isochrones for Vel OB1 and Bochum 7 are plotted assuming the true distance moduli of 11.3 and 13.5~mag, mean $E(V-I_{\rm C})$ equal to 0.46 and 0.91~mag, respectively. In addition, we adopted $R_{\rm V}=$ 3.5, resulting in $A_{\rm V}=$ 1.16~mag for Vel OB1 and 2.28~mag for Bochum 7. The isochrones are from \citet{bert1994} and are plotted for the age of 6~Myr for Vel OB1 (long-dashed line) and 4, 10 and 20 Myr for Bochum 7 (continuous lines). The short-dashed lines are the pre-main sequence isochrones from \citet*{sies2000} for the same three ages as those of \citet{bert1994}. All isochrones are for solar metallicity. In addition, for both stellar systems we show the boundaries of $\delta$~Scuti instability strip after \citet{pamy2000}.} 
\label{cmd}
\end{figure*}

The number of identified members of Vel OB1 and Bochum 7 in our field is too small to try an independent derivation of other parameters than the mean reddenings given above. The remaining parameters are therefore adopted from the previous studies of much larger samples of Vel OB1 and Bochum 7 members, especially those of \citet{sung1999} and \citet{cort2003}. In particular, we adopted the true distance moduli of 11.3 and 13.5~mag for Vel OB1 and Bochum 7, respectively. In addition, \citet{sung1999} and \citet*{cort2008} suggested higher-than-average $R_{\rm V} = A_{\rm V}/E(B-V)$ ratio. Adopting $R_{\rm V}=$ 3.5, we obtained apparent distance moduli of about 12.5~mag for Vel OB1 and 15.8~mag for Bochum 7. 
These distance moduli and the reddenings derived above were used to plot the isochrones in the colour-magnitude diagram (Fig.~\ref{cmd}). As can be seen from this figure, the bright members of Vel OB1 fit quite well all three isochrones we show (for the age of 4, 10 and 20 Myr). In particular, ALS\,1135 fits very well the 4-Myr isochrone, close to the age derived in Sect.~\ref{age} from stellar parameters of the primary. The scatter of the location of certain members should be interpreted rather in terms of differential reddening than age spread. Note that for Vel OB1 the isochrone for 6~Myr adopted after \citet{sung1999} fits the location of the two bright Vel OB1 members quite well.

Using Fig.~\ref{cmd} we can also discuss the membership of variable stars, especially pulsating ones. Given the strong contamination by field stars in our field of view and crossing of isochrones for Vel OB1 and Bochum 7, our photometric diagrams (Fig.~\ref{color_color} and \ref{cmd}) allow an unambiguous determination of membership only for the brightest stars. In particular, V1 (ALS\,1135) and V4 are members of Bochum 7, V3 is a field star and V2 is a certain member of Vel OB1. In view of the detected two periodicities in V2, this star can be classified as an SPB pulsator. Three $\delta$~Scuti stars (V5, V6 and V7) may be members of Vel OB1. 
The possibility that they are field stars, however, cannot be excluded. Of the remaining three $\delta$~Scuti stars, V10 and V12 fall into $\delta$~Scuti instability strip of Bochum 7. If the association is indeed only 4 Myr old, the two $\delta$~Scuti stars would be pre-main sequence (PMS) objects as can be judged from the PMS isochrones plotted in Fig.~\ref{cmd}. An additional argument in favour of the PMS status of V10 and V12 is their location in the two-colour diagram (Fig.~\ref{color_color}). They show a 0.2--0.3~mag excess in $(U-B)$, an indication of the presence of an accretion disk (\citealt{rebu2000,rebu2002}, \citealt*{delg2011}). V15 and V17 are clearly field objects. The membership of the remaining variable stars cannot be decided using our photometric diagrams.

\section{Summary and discussion}
\label{sconc}
The present study was devoted mainly to the massive eclipsing binary system ALS\,1135, a member of the distant OB association Bochum 7. The analysis of the new photometric observations of ALS\,1135 revealed that periodic variations with frequency 2.31095~d$^{-1}$ found in this star from the ASAS-3 data \citep{pimi2007} were caused by a contamination by the neighbouring eclipsing system V4. The new photometry and spectroscopy of ALS\,1135 allowed us to derive accurate parameters of the system, including masses and radii which were obtained with an accuracy better than 1 and 3\%, respectively. 
In comparison with previous results, we obtained a larger value of the orbital inclination and, as a consequence, smaller masses and radii of the components. The masses and radii we found were compared with those from stellar evolutionary models. In this way we estimated that the age of the system amounts to 4.3 $\pm$ 0.5~Myr. For obvious reasons, the parameters of the B-type secondary do not constrain the age so tightly as those of the primary, but are consistent with the above value. The age of ALS\,1135 is also in accordance with the location of the members of Bochum 7 on the isochrone (Fig.~\ref{cmd}).

The presence of O6 and O7-type stars in Bochum~7 implies a very young age of this association, a few Myr at most. The ages of such young stellar systems are usually poorly constrained from isochrone fitting. This is because the result can be affected by a high and variable reddening across a young stellar system, low number of stars close to the turn-off point, inadequacies of transformation from observed (magnitudes, colour indices) to theoretical (absolute magnitude, effective temperature) parameters and the necessity of simultaneous fitting of mean reddening and distance modulus. The method was used by \citet{sung1999} who estimated the age of Bochum 7 for 6~Myr with no significant age spread. 
An age range of 2--7 Myr was derived by \citet{arco2007}\footnote{\citet{arco2007} cite the paper by \citet{cort2003} as the source of this age estimate, but this is not the case. Possibly the other reference (Corti, 2005, Ph.\,D.\,Thesis) is the correct source of this result.}. On the other hand, \citet{crow2006} gives 2.8 $\pm$ 0.5~Myr, but this value was a mean of estimated ages of three O-type stars that belong to Bochum 7 (ALS\,1131, 1135, and 1145). The ages of these stars were assumed to depend solely on their spectral type and luminosity class and therefore the quoted uncertainty is rather underestimated. The age we derived (4.3 $\pm$ 0.5 Myr) is fairly consistent with the previous determinations but is much more precise. It is therefore clear that massive eclipsing binaries can be used to derive ages of open clusters and OB associations with a much better precision than can be achieved from isochrone fitting.  

As a by-product of the study of ALS\,1135 we discovered 17 variable stars, including SPB star belonging to Vel OB1 association and six $\delta$~Scuti stars. Three of them are probable members of Vel OB1, while the other two, V10 and V12, may belong to Bochum 7. If this is the case, they would be pre-main sequence objects. Future seismic modelling of these stars will benefit from our precise determination of the age of Bochum 7 because precise age will better constrain the models.

\section*{Acknowledgements}
This paper uses observations made at the South African Astronomical Observatory (SAAO). 
We thank anonymous referees for comments that have helped us to improve the paper. We are indebted to Prof.~M.\,Jerzykiewicz for his comments made upon reading the manuscript. We also thank Dr.~Z.~Ko{\l}acz\-kow\-ski for providing us the reduced spectrum of ALS\,1135 taken with the MIKE spectrograph.

This work was supported by the Proyecto FONDECYT No.~3085010 and Polish MNiSzW grant N\,N203\,302635 and NCN grant 2011/01/B/ST9/05448. 
NCAR is partially funded by the National Science Foundation. The research leading to these results has received
funding from the European Community's Seventh Framework Programme (FP7/2007-2013) under grant agreement no. 269194.

\bibliographystyle{mn2e}
\bibliography{ALS1135} 
\bsp
\label{lastpage}
\end{document}